\pdfoutput=1
\documentclass[final,authoryear,3p,times]{elsarticle}

\usepackage[utf8]{inputenc}
\usepackage[T1]{fontenc}
\usepackage{lmodern}
\usepackage{amsmath}
\allowdisplaybreaks 
\usepackage{nomencl}
\usepackage{xltabular}
\usepackage{booktabs}
\usepackage{graphicx}
\graphicspath{{figs/}{./}}
\usepackage{subcaption}
\usepackage{booktabs}
\usepackage{threeparttable}
\usepackage{threeparttablex}
\usepackage{tablefootnote}
\usepackage{siunitx}
\usepackage{amsmath,amssymb,amsthm,mathtools}
\usepackage{algorithm}
\usepackage{algpseudocode}
\usepackage{enumitem}
\usepackage{pdflscape}
\usepackage{appendix}
\usepackage{url}
\usepackage{float}
\usepackage{xcolor}
\usepackage[normalem]{ulem}

\usepackage{multirow}
\usepackage{tabularx}
\usepackage{array}
\usepackage{longtable}

\usepackage{natbib}
\usepackage{orcidlink} 

\usepackage{hyperref}
\usepackage[nameinlink,capitalize,noabbrev]{cleveref}

\hypersetup{
	colorlinks,
	citecolor=blue,
	filecolor=blue,
	linkcolor=blue,
	urlcolor=blue
}

\journal{Energy Economics}

\begin{document}
	
\begin{frontmatter}
	
	\title{Optimal power expansion planning under uncertainty in sub-Saharan Africa: A stochastic dual dynamic programming approach}
	
	\author{Arega Getaneh Abate\orcidlink{0000-0002-5517-2585}\corref{cor1}}
	\ead{ageab@dtu.dk}

	\author{Xiao-Bing Zhang\orcidlink{0000-0001-8155-5769}}
	\ead{xiazhan@dtu.dk}
	\author{Xiufeng Liu\orcidlink{0000-0001-5133-6688}}
	\ead{xiuli@dtu.dk}
	\author{Ruyu Liu\orcidlink{0000-0003-2130-9122}}
	\ead{ruyli@dtu.dk}
	\author{Per Sieverts Nielsen\orcidlink{0000-0003-4090-7400}} 
	\ead{pernn@dtu.dk}
	
	\cortext[cor1]{Corresponding author}
	
	\address[DTU]{Department of Technology, Management and Economics, Technical University of Denmark, Kgs. Lyngby, Denmark}

\begin{abstract}
\noindent
Universal electricity access and decarbonization are interlinked challenges for sub-Saharan Africa. Generation expansion planning methods used in the region rarely represent sequential and irreversible investment under demand and hydrological uncertainty. This paper develops a multistage stochastic generation expansion model solved with stochastic dual dynamic programming (SDDP) to address the problem. The model represents demand and country-specific Markov hydrological uncertainty, accounts for technology-specific construction lead times, and endogenous reliability. The model is applied to the Ethiopia and Zimbabwe power systems over 2025--2050. The results show that for Ethiopia, the least-cost pathway is hydropower, with wind and solar sources providing increasing diversification. The model reduces cumulative expected unserved energy by about 78 TWh relative to a lower-cost deterministic expected value investment plan. For Zimbabwe, the least-cost pathway expands solar, hydropower, and gas while retaining residual coal as inherited coal capacity retires. Additionally, carbon pricing reduces emissions and increases resource costs by about 3\% to 5\%, creating a gradual cost-emissions trade-off. Combining a carbon tax policy with concessional solar financing renders the lowest average annual emissions while keeping resource costs below the baseline, which solves the model without emission constraint. At approximately matched expected emissions, emission caps have higher net resource costs than the corresponding carbon tax scenarios. The overall results imply that adaptive investment model can reduce power shortage exposure, while carbon policy design and lower clean investment costs shape the cost of decarbonization in developing power systems.
\end{abstract}

		
\end{frontmatter}

\section{Introduction}\label{sec_intro}

Sub-Saharan Africa faces the dual challenge of expanding electricity access and reducing power sector emissions. Nearly 600 million people on the continent lack access to electricity, with the deficit most acute in rural and peri-urban areas of low-income countries \citep{IEA2023Africa}. More specifically, urbanization, industrial development, and broader electrification drive demand growth across the region \citep{IEA2023Africa,Mensah2024}. Meeting this demand while expanding access requires sustained investment in generation infrastructure over several decades. To this end, generation expansion decisions involve demand and hydrological uncertainty, particularly in settings where access to private capital is limited and national planning authorities manage capacity expansion \citep{Bazilian2012}.

Power system generation expansion planning (GEP) in the developing countries' contexts differs fundamentally from planning in liberalized electricity markets. Unlike market-based power systems, in Africa wholesale electricity markets are often absent. In these countries, retail tariffs are administratively set and frequently fail to recover full costs. Such independent power producers operate through negotiated bilateral contracts as there are no short-run spot markets and the market is not liberalized. In addition, project finance also faces constraints from sovereign risk, currency exposure, and limited domestic capital depth \citep{Steinbuks2010,Falchetta2021}. In most developing countries, the primary decision-maker is usually a national utility where it solely determines the timing and scale of generation capacity expansions to satisfy the growing demand and carbon policy constraints at minimum social cost \citep{Blimpo2024}. However, generation expansion planning is a long-term public investment problem that integrates cost, resource adequacy, emissions, and uncertainty.

Long-term power sector investment is both dynamic and partly irreversible. Capacity committed today affects the cost and feasibility of all future investment decisions. Demand evolves with population growth, electrification, and climate-related changes in cooling or irrigation loads \citep{Dagnachew2023}. These drivers resist accurate forecasting over a 25-year planning horizon, leaving uncertainty that creates value in adapting future decisions as new information emerges \citep{Dixit1994,Fleten2004}. Stochastic dual dynamic programming provides a computational approach for such multistage problems \citep{Pereira1991}. The method approximates downstream value functions with piecewise-linear cuts generated from simulated states and dual information \citep{PhilpottGuan2008,Guigues2017,Dowson2021}. The model is widely used in stochastic hydrothermal scheduling \citep{deQueiroz2016} and has recently been applied to long-term generation expansion problems \citep{HolePhilpottDowson2024,Sun2025}.

The energy transition adds a strategic policy layer to this investment problem. Decarbonization can be pursued through price-based instruments such as carbon taxes and quantity-based instruments such as emissions caps. While the relative performance of these instruments under uncertainty is a classic economic topic, in long-term generation expansion planning, carbon policy operates within an evolving system shaped by investment timing and resource adequacy \citep{Weitzman1974}. A low carbon price may have little effect on investment or/and dispatch, while greater stringency can accelerate the retirement of thermal generation before sufficient low-carbon capacity is available. Annual emissions caps can impose similar constraints when low-carbon alternatives are limited during periods of high demand or/and low renewable availability \citep{Blyth2007,Fuss2008,GoulderParry2008}. These considerations are vital for developing power systems where emissions reduction must proceed alongside substantial generation capacity expansion \citep{Dagnachew2023,IMFSSAClimate2024}. For this reason, carbon policy can be a part of a broader planning problem involving investment, reliability, and uncertainty.

The main analysis in this paper considers two structurally different countries with nationally planned power sectors. Ethiopia possesses one of Africa's largest hydropower resource endowments, and more than 90\% of its installed capacity is hydro-based \citep{IEA2023Africa,EEO2025}. Although the Grand Ethiopian Renaissance Dam (GERD) increased the existing capacity, hydrological variability remains a primary source of supply risk in a country where drought can recur periodically \citep{Tekleab2020}. To tackle this challenge, Ethiopia needs to expand generation to support demand growth while managing drought exposure and continued dependence on hydropower. On the other hand, Zimbabwe faces a different challenge in its power system expansion planning. Its power system relies on the Kariba South hydropower complex, aging coal-fired generation at Hwange and gas-fired capacity. Available capacity is often constrained by deferred maintenance, coal-supply disruptions, in addition to drought-related reductions in water levels \citep{Spalding2017}. Zimbabwe also holds emissions reduction commitments under the Paris Agreement as it has emitting generation mix relatively. Considering this, its planning problem needs to combine resource adequacy, the retirement of existing thermal capacity, and new low-carbon investment.

This paper addresses three research questions. \textit{First}, how can adaptive generation expansion planning represent sequential demand and hydrological uncertainty? \textit{Second}, how do these uncertainties affect least-cost investment pathways in different power systems, and what are the value of adaptive planning and the value of perfect information? \textit{Third}, at approximately matched expected emissions, how do price- and quantity-based carbon instruments differ in resource cost and expected unserved energy when technology costs and resource limits constrain the transition?

While addressing these research questions, the paper makes four main contributions. First, it develops a multistage stochastic generation expansion model in which demand and hydrological conditions are revealed sequentially, hydrology follows country-specific Markov dynamics, and investment adapts to available information. The model is a continuous multistage stochastic linear program solved with stochastic dual dynamic programming (SDDP). Second, the paper quantifies the value of adaptive investment relative to deterministic and perfect information benchmarks. In both applications, the main benefit of the proposed model arises from lower exposure to supply shortages under demand and hydrological uncertainty. Third, the analysis separates real resource costs from domestic carbon-tax transfers and compares price- and quantity-based carbon instruments at approximately matched expected emissions. In Zimbabwe, emissions caps carry modestly higher net resource costs than the corresponding carbon taxes. Fourth, the model is applied to Ethiopia and Zimbabwe over 2025--2050 using country-specific generation fleets, demand trajectories, and hydrological calibration using real data. The comparison shows how differences in resource endowments and inherited fleets shape adaptive investment pathways, providing a stochastic counterpart to existing African planning studies \citep{Juturu2023,Gebremeskel2023}.

The remainder of the paper is organized as follows. Section \ref{sec_literature} reviews the related literature. Section \ref{sec_methodology} presents the multistage stochastic GEP model and its SDDP solution. Section \ref{sec_data} describes data inputs, scenario design, and calibration. Section \ref{sec_results} presents results. Section \ref{sec_discussion} discusses the findings and policy implications. Finally, Section \ref{sec_conclusion} concludes the paper.
	
\section{Related literature}\label{sec_literature}

Generation expansion planning has a long history in operations research and energy economics. Early studies formulated the problem as a deterministic dynamic optimization and showed that the timing of capacity expansion affects system cost and feasibility \citep{Petersen1973,Masse1959}. Later mixed-integer linear programming models allowed more detailed representation of technology choice and operating constraints and became widely used in vertically integrated utility planning \citep{KoltsaklisDagoumas2018,Pfenninger2014}. Electricity market liberalization subsequently shifted part of the literature toward generation expansion investment under market uncertainty \citep{Murphy2005,Fleten2004}. However, this strand is less directly applicable to many sub-Saharan African power systems, where wholesale markets remain limited and capacity expansion is often coordinated by national utilities or planning institutions. In these countries, social-planner GEP models evaluate long-term investment by minimizing expected discounted social cost subject to energy balance, capacity, and resource-adequacy constraints \citep{Bazilian2012,Blimpo2024}.

From modeling perspectives, stochastic programming extends GEP by allowing investment decisions to respond to uncertainty revealed over time \citep{ShapiroDentchevaRuszczynski2009,Birge1997}. As a solution method, \citet{Pereira1991} introduce SDDP for large multistage stochastic linear programs, using dual information to construct piecewise-linear approximations of downstream value functions. Its convergence and statistical properties have since been studied extensively \citep{PhilpottGuan2008,Shapiro2011,Guigues2017}, while current implementations support extensions including discrete decisions, risk averse behaviors, and Markovian uncertainty \citep{Dowson2021}. Specifically, \citet{deQueiroz2016} analyzes the use of SDDP in stochastic hydro-thermal scheduling, where sequential hydrological uncertainty is central to operating decisions. \citet{Zou2019} develop stochastic dual dynamic integer programming for multistage unit commitment, extending the approach to problems with discrete decisions. \citet{BotterudKorpas2007} study stochastic generation investment timing for a decentralized investor facing load-growth and electricity price uncertainty in a restructured Nordic market. \citet{HolePhilpottDowson2024} jointly model capacity investment and operational dispatch with SDDP in the hydro-dominated New Zealand system. More recently, \citet{Sun2025} use an SDDP-based generation expansion model to compare carbon taxes and emissions caps over a long planning horizon and show that the two instruments can lead to different investment pathways.

While considering GEP, the design of carbon policy has been studied extensively in environmental and energy economics. \citet{Weitzman1974} establishes that the relative performance of price- and quantity-based instruments under uncertainty depends on the slopes of marginal benefits and marginal costs. \citet{Kolstad1996} further shows that price instruments tend to perform better when abatement cost uncertainty is large and marginal damages are relatively flat, whereas quantity instruments become more attractive when uncertainty about damages is more important. In practice, instrument choice also depends on implementation costs, distributional effects, and political economy considerations \citep{GoulderParry2008}. In the electricity sector, \citet{HeWangWang2012} compare cap-and-trade and carbon taxes in a game-theoretic generation expansion model and document differences in emissions, investment, and welfare across the two instruments. Carbon policy uncertainty also interacts with irreversible investment decisions and can delay low-carbon investment by increasing the value of waiting \citep{Blyth2007,Fuss2008}.

For developing country power systems, evidence on carbon policy design is more limited. \citet{IMFSSAClimate2024} and \citet{WorldBankPower2024} examine how carbon finance and investment conditions affect low-carbon power sector development, while \citet{Dagnachew2023} show that productive uses of electricity need to be represented in African energy planning to avoid understating future investment requirements. The effectiveness of carbon pricing in these systems depends on the level of the carbon price as well as the availability and cost of low-carbon technologies, financing conditions, and the persistence of the policy framework. Less attention has been given to settings in which technology costs, existing assets, and resource constraints already push the least-cost expansion pathway toward lower-carbon generation. In such systems, the additional effect of carbon policy may differ from cases in which the policy itself is the main driver of technology choice.

The empirical literature on electricity supply in sub-Saharan Africa examines access, investment, reliability, and institutional constraints from several perspectives. \citet{Bazilian2012} analyze the electricity access challenge and emphasize the importance of market and institutional barriers, while \citet{Trotter2017} review electricity planning and implementation across the region. \citet{Egli2023} estimate that universal household electrification in sub-Saharan Africa would require investment well above current financing flows. \citet{Mensah2024} provides evidence on the economic consequences of electricity shortages, while \citet{Steinbuks2010} discuss the use of self-generation in response to unreliable supply and \citet{Farquharson2018} quantify the associated cost and emissions of backup generation. For Ethiopia, \citet{Juturu2023} examine the interaction among electrification, groundwater irrigation, and generation expansion. \citet{Gebremeskel2021} analyze long-term electricity demand, while \citet{Gebremeskel2023} study least-cost generation expansion using a deterministic optimization framework. \citet{Tekleab2020} analyze the water balance of Upper Blue Nile catchments, the basin on which much of Ethiopia's hydropower depends. For Zimbabwe and the wider Southern African region, \citet{Spalding2017} examine climate-related reductions in hydropower availability across the Southern African Power Pool and the Zambezi basin. Similarly, declining solar investment costs have improved the competitiveness of solar PV \citep{IRENA2023}, while regional electricity pooling can reduce some of the costs associated with hydrological variability \citep{Bissiri2024}. Financing remains an important constraint, with recent assessments emphasizing the role of concessional and climate finance in supporting large-scale renewable investment \citep{IMFSSAClimate2024}.
	
However, three gaps in the current literature are relevant to this paper. \textit{First}, closely related multistage stochastic models address important elements of generation planning under uncertainty, including stochastic hydro operation \citep{HolePhilpottDowson2024}, hydro-thermal scheduling \citep{deQueiroz2016}, and multistage unit commitment \citep{Zou2019}. Less attention has been given to long-horizon expansion models in which demand and hydrological conditions evolve sequentially and investment adapts as their realizations become known. \textit{Second}, studies comparing price- and quantity-based carbon emission instruments under endogenous technology choice \citep{HeWangWang2012,Sun2025} provide limited evidence for nationally planned systems in which inherited assets, technology costs, resource constraints, and hydrological uncertainty shape the response to carbon policy. \textit{Third}, country level planning studies for sub-Saharan Africa \citep{Juturu2023,Gebremeskel2023} rely largely on deterministic formulations and provide limited evidence on how sequential demand and hydrological uncertainty affect adaptive investment pathways and the value of adaptive planning. This paper addresses these gaps through a multistage stochastic generation expansion model applied to two structurally different sub-Saharan African power systems. The model represents demand uncertainty and country-specific Markov hydrological dynamics with adaptive investment, and it quantifies the value of adaptive planning relative to deterministic and perfect information benchmarks. The Zimbabwe application compares carbon policy instruments at approximately matched expected emissions. Table \ref{tab_literature_positioning} summarizes the positioning relative to selected closely related studies.

\begin{table}[H]
	\centering
	\caption{Positioning of the paper relative to selected closely related studies. A \checkmark denotes substantive treatment of the feature, while a dash indicates that the feature is not a central element of the study. \label{tab_literature_positioning}}
	\begin{threeparttable}
		\small
			\begin{tabular}{lccccc}
				\toprule
				Paper & Demand & Hydrology & Adaptive & Carbon & Developing\\
				& uncertainty & uncertainty & investment & policy & country\\
				\midrule
				\citet{deQueiroz2016}            & --         & \checkmark & --         & --         & -- \\
				\citet{Zou2019}                  & \checkmark & --         & --         & --         & -- \\
				\citet{BotterudKorpas2007}       & \checkmark & --         & \checkmark & --         & -- \\
				\citet{HolePhilpottDowson2024}   & --         & \checkmark & --         & --         & -- \\
				\citet{Sun2025}                  & \checkmark & --         & \checkmark & \checkmark & -- \\
				\citet{HeWangWang2012}           & --         & --         & --         & \checkmark & -- \\
				\citet{Bissiri2024}              & --         & \checkmark & --         & --         & \checkmark \\
				\citet{Juturu2023}               & --         & --         & --         & --         & \checkmark \\
				\citet{Gebremeskel2023}          & --         & --         & --         & --         & \checkmark \\
				\citet{Dagnachew2023}            & --         & --         & --         & --         & \checkmark \\
				\textbf{This paper}              & \checkmark & \checkmark & \checkmark & \checkmark & \checkmark \\
				\bottomrule
			\end{tabular}
	\end{threeparttable}
\end{table}

\section{Methodology}\label{sec_methodology}

Long-term generation expansion planning can be formulated as a static problem that identifies a least-cost generation capacity mix or as a dynamic problem that determines investment and operating decisions over time. A dynamic formulation becomes important when demand evolves, investment commitments are difficult to reverse, and insufficient generation capacity and shortage in energy supply carries an economic cost. When demand and hydrological conditions are uncertain, investment decisions must also account for how current choices affect the feasible and economically cost-efficient options available in later years. This problem can be cast as a multistage stochastic program represented through a dynamic programming recursion \citep{Bellman1952}. However, explicit enumeration of the corresponding scenario tree becomes impractical over a long horizon as the number of possible histories grows rapidly with the number of stages and uncertainty realizations.

We solve this multistage stochastic linear GEP using stochastic dual dynamic programming (SDDP) \citep{Pereira1991}. As a solution method, SDDP avoids enumeration of the full scenario tree by approximating future cost functions with piecewise-affine cuts generated from dual information. Its forward simulations identify states visited by the current policy, while the backward passes refine the continuation value approximations at those states. Under the standard assumptions for multistage stochastic linear programs, these approximations converge to the optimal value function \citep{PhilpottGuan2008,Shapiro2011,Guigues2017}. In SDDP, the policy is adaptive and nonanticipative where investment and operating decisions at each stage depend only on information available at that time. The information includes installed capacity, previous investment commitments, realized demand, and the current hydrological condition. By contrast, a deterministic formulation fixes future investment decisions to a prescribed trajectory, while separate deterministic scenario runs optimize different assumed futures independently. Neither produces a common policy that adjusts investment as uncertainty is realized. This distinction is important for the two power systems considered in this paper as demand and hydrological conditions affect investment timing, capacity requirements, and exposure to supply shortages. More specifically, the model represents two exogenous sources of uncertainty, annual electricity demand and hydrological conditions. Their persistence is reflected in the information available to future decisions. To this end, demand is represented by a deterministic central trajectory $\bar{\phi}_t$ and a stochastic deviation $u_t$ following
\begin{equation}\label{eq_demand_process_refined}
	u_{t+1}=\rho\,u_t+\sigma_t\,\varepsilon_{t+1},\qquad
	\varepsilon_{t+1}\sim\mathcal{N}(0,1), \qquad \phi_t=\bar{\phi}_t+u_t,
\end{equation}
where $\phi_t$ denotes realized annual electricity demand in year $t$. The parameter $\rho$ governs the persistence of demand deviations, while $\sigma_t=\sigma_0\bar{\phi}_t$ scales the innovation with the central demand ($\bar{\phi}_t$) trajectory to keep relative uncertainty approximately stable as demand grows. To account for demand uncertainty, the deviation $u_t$ enters the state representation, allowing the continuation value to depend on the current demand realization and its implications for future demand.
	
\renewcommand{\arraystretch}{1.12}

\begin{xltabular}{\textwidth}{@{}lX@{}}
	\caption{Main notation used throughout the paper.}\label{tab_notation}\\
	
	\toprule
	\textbf{Symbol} & \textbf{Description}\\
	\midrule
	\endfirsthead
	
	\multicolumn{2}{@{}l}{\itshape\small
		\tablename \thetable{} (continued)}\\
	\toprule
	\textbf{Symbol} & \textbf{Description}\\
	\midrule
	\endhead
	
	\midrule
	\multicolumn{2}{r@{}}{\itshape\small Continued on next page}\\
	\endfoot
	
	\bottomrule
	\endlastfoot
	
	\multicolumn{2}{@{}l}{\textbf{Indices and sets}}\\
	$t\in\{1,\dots,T\}$ & Planning year (SDDP stage)\\
	$e\in\mathcal{E}$ & Generation technology\\
	$k\in\{1,\dots,\ell_e-1\}$ & Construction-pipeline bucket ($k$ years before commissioning)\\
	$\mathrm{hy}\in\mathcal{E}$ & Reservoir-hydropower technology index\\[2pt]

	\multicolumn{2}{@{}l}{\textbf{State variables (start of year $t$)}}\\
	$X_{te}\ge 0$ & Installed capacity (MW)\\
	$Q_{tek}\ge 0$ & Committed capacity $k$ years from commissioning (MW)\\
	$S_{te}\ge 0$ & Cumulative retired capacity (MW)\\
	$u_t$ & Demand-deviation state (MWh)\\
	$h_t\in\mathcal{H}=\{L,N,H\}$ & Annual hydrological condition, Markov policy-graph node (low-, normal-, high-inflow)\\[2pt]

	\multicolumn{2}{@{}l}{\textbf{Decision variables (year $t$)}}\\
	$G_{te}\ge 0$ & Annual generation (MWh)\\
	$I_{te}\ge 0$ & New capacity committed (MW)\\
	$R_{te}\ge 0$ & Capacity retired (MW)\\
	$Z_{t}\ge 0$ & Annual unserved energy (MWh)\\[2pt]

	\multicolumn{2}{@{}l}{\textbf{Parameters and stochastic inputs}}\\
	$\bar{\phi}_t,\ \phi_t$ & Central and realized annual demand (MWh)\\
	$\rho,\ \sigma_t$ & Demand persistence and innovation scale\\
	$\varepsilon_t$ & Standard-normal demand innovation\\
	$P^{H},\ \mu_h$ & Hydrological transition matrix and condition multiplier\\
	$f_e$ & Technology-specific capacity factor\\
$\eta_e$ & Annual energy-availability coefficient (MWh/MW-year)\\
	$\ell_e$ & Construction lead time (years)\\
	$c_{1e},\ c_{2te},\ c_{4e}$ & Variable (\$/MWh), investment (\$/MW), and fixed O\&M (\$/MW-year) costs\\
	$c_3$ & Value of lost load (\$/MWh)\\
	$r,\ \delta_t$ & Social discount rate and discount factor $\delta_t=(1+r)^{-(t-1)}$\\
	$\varsigma$ & Terminal salvage fraction of last-period investment cost\\
	$\overline{X}_e,\ F_{te}$ & Capacity ceiling and cumulative retirement floor (MW)\\
	$\tau,\ EM_e$ & Carbon tax (\$/kg) and emissions intensity (kg/MWh)\\
	$E^{\max}_t$ & Annual emissions cap (kg)\\
	$X^{\mathrm{init}}_e$ & Initial installed capacity (MW)\\
$Q^{\mathrm{init}}_{ek}$ & Initial construction pipeline (MW)\\
	
\end{xltabular}

To consider hydrological uncertainty, we assume that hydrological conditions follow a three-state annual Markov chain with low-inflow, normal-inflow, and high-inflow states and transition matrix $P^{H}$. Annual hydrological conditions are defined from the historical distribution of annual river inflow. Annual inflows are classified using the one-third and two-thirds empirical quantiles of this distribution, approximately the 33rd and 67th percentiles. Observations below the lower cutoff correspond to the low-inflow condition, observations between the two cutoffs to the normal-inflow condition, and observations above the upper cutoff to the high-inflow condition. This equal-frequency classification provides a parsimonious representation of hydrological variability while maintaining a balanced representation of the historical observations across the three conditions. The multiplier $\mu_h$ is the mean normalized annual inflow within condition $h$, and the elements of $P^{H}$ are estimated from observed transitions between consecutive annual conditions. In hydrological condition $h$, annual reservoir-hydropower energy availability is scaled by the multiplier $\mu_h$, with $\mu_L<\mu_N<\mu_H$. The multipliers are normalized such that $\sum_h \pi_h \mu_h=1$, where $\pi$ is the stationary distribution of $P^{H}$, which makes the central hydropower capacity factor the long-run expected availability level. The transition matrix describes year-to-year hydrological persistence. If year $t$ is a low-inflow year, the corresponding row of $P^{H}$ gives the probabilities that year $t+1$ is a low-, normal-, or high-inflow year. Demand evolves through the persistent deviation $u_t$ and the stagewise-independent innovation $\varepsilon_{t+1}$. At the beginning of year $t$, the continuous state consists of installed capacity, the construction pipeline, cumulative retirements, and the demand deviation $u_t$. The current hydrological condition $h_t$ is represented separately by the corresponding node of the SDDP policy graph. It affects the stage problem through the condition-dependent multiplier $\mu_{h_t}$ and evolves according to $P^{H}$.

\subsection{Model formulation}\label{sec_methodology_formulation}

The national generation expansion problem is formulated as a multistage stochastic program over years $t\in\mathcal{T}=\{1,\dots,T\}$ and technologies $e\in\mathcal{E}$. Each stage represents one planning year, and all physical and economic decisions are annual. The same core formulation is applied to Ethiopia and Zimbabwe, with carbon policy instruments added for the Zimbabwe analysis. Table \ref{tab_notation} summarizes the notation. State variables are observed at the beginning of each year, and decisions are made after that year's demand and hydrological condition are realized. All decision variables are continuous and each stage problem is linear.
The social accounting stage cost $C_t$ combines investment, fixed O\&M, and variable operating costs with the economic cost of supply inadequacy. Unserved energy is valued at the value of lost load,
\begin{equation}\label{eq_stage_cost_refined}
	C_t = \sum_{e} c_{2te} I_{te} + \sum_{e} c_{4e} X_{te}
	+ \sum_{e} c_{1e} G_{te} + c_3 Z_{t} .
\end{equation}

The planning stage cost adds carbon-tax payments to the social accounting cost,
\begin{equation}\label{eq_planning_cost}
\widetilde{C}_t = C_t + \tau\sum_{e} EM_e\,G_{te},
\end{equation}
where $\tau=0$ in scenarios without a carbon tax, where a single objective function covers every scenario.
Different cost measures are presented in the analysis. The \emph{gross resource cost} is the discounted sum of investment, fixed O\&M, and variable operating costs, and the \emph{net resource cost} deducts the terminal salvage value from it. The \emph{total expected cost} adds the discounted value of unserved energy to the net resource cost. The \emph{cash cost} adds carbon-tax payments to the gross resource cost. The carbon-tax payments enter the planning objective function through $\widetilde{C}_t$ in \eqref{eq_planning_cost}, where they affect generation and investment decisions, and they are treated as domestic transfers that are excluded from the resource cost and the total expected cost. This treatment separates the incentive effect of carbon pricing from its treatment in social accounting and allows comparison with the quantity-based carbon instrument on a common resource cost basis.

The objective function in \eqref{eq_obj} also includes a terminal salvage value for new capacity remaining in service beyond the planning horizon. Eligible capacity is defined as $X_{T+1e}-\bigl(X^{\mathrm{init}}_{e}-S_{T+1e}\bigr)$,
which measures end-of-horizon installed capacity net of the surviving inherited fleet. The states $X_{T+1e}$ and $S_{T+1e}$ follow from the commissioning and cumulative retirement transitions in \eqref{eq_commission} and \eqref{eq_retire}, evaluated at $t=T$. Salvage value is assigned only to capacity added during the planning horizon as investment in the inherited fleet is a sunk cost. This treatment accounts for the residual value of new investments that remain productive beyond the model horizon while preserving retirement incentives for inherited capacity. The terminal value also limits finite-horizon bias against long-lived technologies whose useful lives extend beyond 2050. 

The timing within each stage is as follows. One SDDP stage corresponds to one year. At the start of year $t$, the planner observes the realized annual demand $\phi_t=\bar{\phi}_t+u_t$ and the annual hydrological condition $h_t$. Annual generation, unserved energy, investment, and retirement are then chosen for that year. Committed investment enters service after the technology-specific construction lead time $\ell_e$. Hydrology affects the stage only through the annual hydropower energy available, which is scaled by $\mu_{h_t}$, and the condition evolves to the next year according to the Markov matrix $P^{H}$. A low-inflow year lowers hydropower energy availability through $\mu_L<1$, raising reliance on other technologies or on unserved energy, and its persistence carries into the following year through $h_{t+1}$. Retirement $R_{te}$ enters the next-year capacity $X_{t+1e}$, where it acts at the end of year $t$ while generation in year $t$ uses the capacity $X_{te}$ available during that year. Assume $\delta_t=(1+r)^{-(t-1)}$ denotes the discount factor associated with the social discount rate $r$, and let $\mathrm{hy}$ index reservoir hydropower. The multistage stochastic program is

\begin{subequations}\label{eq_model}
	\begin{align}
		\min \ & \mathbb{E}\Big[\textstyle\sum_{t=1}^{T} \delta_t\, \widetilde{C}_t
		- \delta_T\,\varsigma\sum_{e} c_{2Te}\, \big(X_{T+1e}-X^{\mathrm{init}}_{e}+S_{T+1e}\big) \Big]
		\label{eq_obj}\\[2pt]
			&\text{s.t.}\nonumber\\
			& \textstyle\sum_{e} G_{te} + Z_{t}
			= \bar{\phi}_t+u_t,
			&& \forall t
			\label{eq_balance}\\
			& G_{te} \le \eta_e\,X_{te},
			&& \forall t,\ e\neq\mathrm{hy}
			\label{eq_avail}\\
			& G_{t,\mathrm{hy}} \le
			\eta_{\mathrm{hy}}\,\mu_{h_t}\,X_{t,\mathrm{hy}},
			&& \forall t
			\label{eq_hydroavail}\\
			& X_{t+1e}
			= X_{te} + I_{te}\,\mathbf{1}\{\ell_e=1\}
			+ Q_{te1}\,\mathbf{1}\{\ell_e>1\} - R_{te},
			&& \forall t,\ e
			\label{eq_commission}\\
			& Q_{t+1ek} =
			\begin{cases}
				Q_{tek+1}, & k=1,\dots,\ell_e-2,\\
				I_{te},     & k=\ell_e-1,
			\end{cases}
			&& \forall t<T,\ e:\ell_e>1
			\label{eq_pipeline}\\
			& S_{t+1e} = S_{te}+R_{te},
			&& \forall t,\ e
			\label{eq_retire}\\
			& R_{te}\le X^{\mathrm{init}}_e-S_{te},
			&& \forall t,\ e
			\label{eq_retirecap}\\
			& S_{te}+R_{te}\ge F_{te},
			&& \forall t,\ e
			\label{eq_bounds}\\
			& X_{te}\le \overline{X}_e,
			&& \forall t,\ e
			\label{eq_capbound}\\
			& u_{t+1} = \rho\,u_t+\sigma_t\,\varepsilon_{t+1},
			&& \forall t<T
			\label{eq_transition}\\
			& h_{t+1}\sim P^{H}(\cdot\mid h_t)
			&& \forall t<T.
			\label{eq_hydro}
	\end{align}
\end{subequations}

The initial conditions are $X_{1e}=X^{\mathrm{init}}_e$, $S_{1e}=0$, $u_1=0$, and $Q_{1ek}=Q^{\mathrm{init}}_{ek}$, where $Q^{\mathrm{init}}_{ek}$ is the initial construction pipeline calibrated from projects already committed or under construction and is zero when no such project exists. The initial hydrological condition $h_1$ is fixed at the calibrated pre-horizon condition. All decision variables are non-negative. For technologies with construction lead time $\ell_e>1$, the pipeline contains buckets $k\in\{1,\dots,\ell_e-1\}$, and capacity committed in year $t$ becomes available in year $t+\ell_e$. Technologies with $\ell_e=1$ are commissioned directly through \eqref{eq_commission}.

Constraint \eqref{eq_balance} imposes annual energy balance, where annual generation plus annual unserved energy $Z_t$ equals realized annual demand, and unserved energy incurs the value of lost load $c_3$. Constraint \eqref{eq_avail} converts installed capacity into feasible annual energy for each non-hydro technology through the annual energy-availability coefficient $\eta_e=8760\,f_e$, where $f_e$ is the technology-specific capacity factor and $\eta_e$ is the annual energy available in MWh from one MW. For variable renewables such as wind and solar, $f_e$ reflects annual resource availability, and for dispatchable technologies such as coal, gas, geothermal, and bioenergy it represents the planning-level annual utilization. Constraint \eqref{eq_hydroavail} limits annual reservoir-hydropower generation, where $\eta_{\mathrm{hy}}=8760\,f_{\mathrm{hy}}$ sets the long-run annual availability from the central hydropower capacity factor $f_{\mathrm{hy}}$ and the annual Markov multiplier $\mu_{h_t}$ raises or lowers it according to the realized inflow condition. This is the principal physical constraint through which hydrological uncertainty affects investment.

Construction dynamics are represented by constraints \eqref{eq_commission} and \eqref{eq_pipeline}. Investment $I_{te}$ enters the construction pipeline based on the technology-specific lead time $\ell_e$ and advances toward commissioning in subsequent years. For technologies with $\ell_e=1$, committed capacity enters the installed-capacity state directly through constraint \eqref{eq_commission}. Constraint \eqref{eq_retire} tracks cumulative retirement of the inherited fleet, and constraint \eqref{eq_retirecap} limits retirement to capacity available at the beginning of the planning horizon. The lower bound $F_{te}$ in constraint \eqref{eq_bounds} imposes exogenous retirement and phase-out requirements, with additional retirement of inherited capacity determined endogenously. The upper bound $\overline{X}_e$ in constraint \eqref{eq_capbound} represents technology-specific resource and deployment limits. Demand uncertainty evolves through the persistent deviation $u_t$ in \eqref{eq_transition}, while the hydrological condition $h_t$ follows the Markov transition matrix $P^{H}$ in \eqref{eq_hydro}.

The Zimbabwe application adds two carbon policy instruments to \eqref{eq_model}. The carbon tax enters through the planning stage cost $\widetilde{C}_t$ with $\tau>0$. That is because the carbon charge affects generation and investment decisions while the associated tax payment is excluded from resource cost and presented separately in the cash cost measure. Under an annual emissions cap, generation must satisfy
\begin{equation}\label{eq_cap}
\sum_{e} EM_e\,G_{te}\le E^{\max}_t
\end{equation}
in each year. For the carbon policy comparison, the price- and quantity-based instruments are calibrated to approximately matched expected emissions. A root-finding procedure adjusts the annual cap, re-solving the model at each trial value, until expected emissions on a dedicated calibration sample are sufficiently close to those produced by the reference carbon tax. The calibrated instruments are then evaluated on an independent out-of-sample set generated with different random seeds. This separation prevents the policy parameters from being tuned to the sample used for the comparison. Instrument performance is evaluated using resource cost, expected unserved energy, and emissions. The allocation is evaluated using consolidated social resource cost, which excludes domestic tax payments as transfers within the national account. Cash cost records these payments separately. This accounting convention preserves the incentive effect of the carbon price while allowing the price- and quantity-based instruments to be compared on a common resource cost basis.

\subsection{Value-function approximation and SDDP implementation} \label{sec_solution}

SDDP decomposes the multistage stochastic program into stagewise linear subproblems linked through approximations of future cost. At each stage, the downstream value function is approximated from below by affine cuts constructed from dual information obtained at trial states. These approximations are refined iteratively as new states are generated during policy simulation. We implement the model in SDDP.jl \citep{Dowson2021} and solve the stage problems with Gurobi. Hydrological persistence is represented through a Markovian policy graph whose nodes correspond to the low-, normal-, and high-inflow conditions. Suppose
\begin{equation}
x_t:=(X_{t\cdot},Q_{t\cdot\cdot},S_{t\cdot},u_t)
\end{equation}
denotes the continuous state at stage $t$. For policy graph node $i$, corresponding to a hydrological condition, assume $V_t^{(i)}(x_t)$ denotes the expected discounted cost from stage $t$ onward under the optimal policy, with $V_{T+1}^{(i)}(\cdot)=0$. At each node, the value function is approximated by the piecewise affine lower bound
\begin{equation}\label{eq_cuts_refined}
	\widehat V_t^{(i)}(x)	= \max_{k\in\mathcal{K}_t^{(i)}}
	\left\{\beta_{tk}+\pi_{tk}^{\top}x \right\},
\end{equation}
where $\mathcal{K}_t^{(i)}$ denotes the set of cuts accumulated at stage $t$ and node $i$. Each cut $(\beta_{tk},\pi_{tk})$ is constructed from dual information generated during the backward pass. At node $i$, the continuation value is the transition-weighted expectation of the successor value functions, $\sum_{j} P^{H}_{ij}\,\mathbb{E}_{\varepsilon}[V_{t+1}^{(j)}(x_{t+1})]$, and each successor value function $V_{t+1}^{(j)}$ is approximated by its own cuts. In each stage problem, the successor continuation variable is constrained by the successor cuts,
\begin{equation}\label{eq_cut_constraints_refined}
\theta_{t+1}^{(j)} \ge	\beta_{t+1k}^{(j)}	+ \bigl(\pi_{t+1k}^{(j)}\bigr)^{\top}x_{t+1}, \qquad \forall k\in\mathcal{K}_{t+1}^{(j)}.
\end{equation}

The node-$i$ continuation entering the stage objective function is $\sum_{j} P^{H}_{ij}\,\theta_{t+1}^{(j)}$. Demand and hydrological uncertainty enter the algorithm through different mechanisms. The demand innovation $\varepsilon_t$ is stagewise independent and is discretized using five-point Gauss-Hermite quadrature\footnote{Gauss-Hermite quadrature is a numerical rule for representing a continuous normal distribution by a small set of discrete outcomes. It approximates the expectation of a function under the normal density by a weighted sum over a few chosen points, called nodes, with each weight acting as the probability attached to its node. Five nodes reproduce the low-order moments of the distribution closely while keeping the number of subproblems solved in each backward pass small.} in the backward pass. Each SDDP iteration consists of a forward and a backward pass, summarized in Algorithm \ref{alg_sddp} in \ref{app_convergence}. The forward pass samples demand innovations and hydrological transitions, generates a policy trajectory, and records the trial states. The backward pass solves the relevant stage problems at these states and adds cuts that refine the approximation of future cost. The first-stage value provides a lower bound on the optimal expected cost.
Convergence is assessed using a statistical optimality gap criterion. After an initial training period, the current policy is evaluated on a fixed set of in-sample paths generated from the same discrete probability model used in the backward pass. These simulations provide a one-sided 95\% upper confidence bound on expected policy cost. Training terminates when the relative difference between this upper bound and the SDDP lower bound remains below a pre-specified tolerance for several consecutive evaluations. The confidence bound is used only as a convergence diagnostic and is not interpreted as providing formal sequential coverage. A fixed random seed is used to make the training procedure reproducible \citep{PhilpottGuan2008,Shapiro2011,Dowson2021}. After training, policy performance is evaluated on 5,000 independently generated out-of-sample paths, with 1,000 paths generated under each of five random seeds.

The stochastic policy is compared with deterministic and perfect information benchmarks following standard stochastic programming practice \citep{Birge1997}. The deterministic benchmark is the conventional expected value problem, obtained by solving the model along the central demand trajectory under expected hydrology. Expected hydrology is stage-specific and conditional on the calibrated pre-horizon hydrological condition. With $h_1$ represented by its state vector, the expected multiplier in year $t$ is $\mathbb{E}[\mu_t\mid h_1]=h_1^\top(P^H)^{t-1}\mu$. Therefore, the deterministic version is an expected value benchmark. The investment schedule is then fixed and evaluated out of sample, while dispatch and retirement decisions remain responsive to realized demand and hydrological conditions. The perfect information benchmark solves each sampled path with full knowledge of future realizations. The value of the stochastic solution measures the expected cost reduction relative to the deterministic benchmark. The expected value of perfect information measures the additional expected cost reduction attainable with full foresight. All three policies are scored on the same common continuous out-of-sample paths under common random numbers, making the value of adaptive planning and the perfect information value estimates paired differences on a common evaluation measure. These paired comparisons use a common set of 200 evaluation paths. The perfect information benchmark solves each path separately to optimality, which is computationally more demanding than evaluating a fixed policy over the larger production sample. The perfect information value estimate is nonnegative by construction. The stochastic policy is not solved to certified optimality, which makes the estimated value of adaptive planning a conservative lower bound and the perfect information value estimate an upper bound. Finally, cost and expected unserved energy are presented separately, given that adaptive investment may reduce shortage exposure even when differences in investment expenditure are relatively small.
	
\section{Data, calibration, and scenario design}\label{sec_data}

\subsection{System data and parameter calibration}

The model combines country-specific power system data with international technology, resource, and cost datasets. The modeled system is the grid-connected national generation system in each country, and off-grid generation is excluded. For Ethiopia, demand is domestic electricity consumption and the energy balance is loss-free, where distribution losses are not added to the consumption. The Ethiopian fleet is a post-GERD 2025 production of 9,757 MW dominated by hydropower (9,213 MW) with 502 MW of wind \citep{EEP2025}, and Koysha (1,800 MW) is represented in the initial construction pipeline instead of considering it as installed capacity, with commissioning at the end of its construction lead time. For Zimbabwe, demand is electricity delivered into the transmission and system-supply boundary, and imports are embedded in that requirement as calibrated. The official Zimbabwean system total is 2,962 MW, composed of coal (1,680 MW), large hydropower (1,050 MW), and other independent power producers (232 MW) \citep{ZWECompact2025}, which the model resolves into hydropower of 1,120 MW (Kariba South and small hydro), coal of 1,680 MW (Hwange, including Units 7 and 8), solar of 70 MW, and bagasse of 92 MW, with gas an investable technology at zero initial capacity. This reconciliation into modeled technologies is presented in Table \ref{tab_tech_params}. The Ethiopian central demand trajectory begins at 25.18 TWh in 2025, reaches the 40 TWh electrification target by 2030, and grows at 3.5\% per year to about 79.6 TWh in 2050 \citep{EEO2025}, while the Zimbabwean trajectory begins at a calibrated system-supply requirement of 11.082 TWh in 2025 \citep{ZERA2024} and grows at 3.5\% per year to about 26.2 TWh in 2050, with its internal generation and net-import components detailed in \ref{app_calibration}. Stochastic deviations around these central paths follow the AR(1) process in \eqref{eq_demand_process_refined}, with persistence and innovation scale estimated from country-specific detrended annual electricity series over 2000--2023 \citep{Ember2025}, giving pooled values $\rho=0.81$ and $\sigma_t=0.077\,\bar{\phi}_t$.

Technology capacity factors, costs, construction lead times, and emission intensities are summarized in Table \ref{tab_tech_params}. Solar capacity factors are derived from country-specific practical photovoltaic potential yields \citep{GlobalSolarAtlas}, giving 0.196 for Ethiopia and 0.203 for Zimbabwe. The central hydropower factors are 0.45 and 0.50 for Ethiopia and Zimbabwe, respectively. Interannual hydrological variability is represented separately through the state multiplier $\mu_h$. The Ethiopian wind factor of 0.34 is an annual planning benchmark. The factors for the dispatchable technologies are annual utilization ceilings. Investment costs follow technology-specific trajectories \citep{IRENA2023}, with solar PV declining from \$900/kW in 2025 to \$400/kW by 2050 in both countries and Ethiopian wind falling by about 31\%, while thermal and hydropower costs change more gradually. Investment is discounted at a 5\% social rate and booked in the year of commitment. Unserved energy is valued at a central VoLL of \$10,000/MWh, with \$5,000 and \$20,000/MWh used as sensitivities, and the central value is a planning benchmark informed by evidence on the economic cost of unreliable electricity supply for firms in the region \citep{Carlsson2020}. The model uses operational emission coefficients of 924 kgCO$_2$/MWh for coal, 630 for gas, 230 for bioenergy, and 50 for geothermal, while hydro, wind, and solar are assigned zero operational emissions, and the bioenergy coefficient reflects the net accounting convention adopted in the model. Each technology also carries an upper bound on cumulative installed capacity, imposed through \eqref{eq_capbound}. These bounds are long-term planning assumptions on the scale of development considered feasible over the planning horizon. They are exogenous planning assumptions informed by the existing system, broad resource availability and endowment, development plans, and technology costs, and not estimates of technical resource potential. The values are presented in \ref{app_calibration}. Complete cost trajectories, fixed operating costs, and a broader comparison with published value-of-lost-load estimates are provided in \ref{app_calibration}.

\begin{table}[htbp]
	\centering
	\caption[Technology parameters and initial installed capacity.]{Technology parameters and initial installed capacity.\tablefootnote{Initial capacities and CAPEX refer to 2025. Complete cost trajectories to 2050 and fixed operating costs are given in \ref{app_calibration}. Solar capacity factors are derived from practical photovoltaic potential yields. Hydropower and Ethiopian wind capacity factors are annual planning benchmarks, while those for coal, gas, geothermal, and bioenergy are annual utilization ceilings. Emission intensities are net planning coefficients, with bioenergy represented on a net-accounting basis. For Zimbabwe, the official installed capacity of 2,962 MW comprises 1,680 MW of coal, 1,050 MW of large hydropower, and 232 MW of other independent power producers. The model reconciles the latter category into small hydropower, solar, and bagasse.}\label{tab_tech_params}}
	\begin{threeparttable}
		\footnotesize
			\begin{tabular*}{\textwidth}{@{\extracolsep{\fill}}lrrrrrr}
				\toprule
				\multicolumn{7}{c}{\textbf{Ethiopia}}\\
				\midrule
				Technology & Initial capacity & CAPEX & Variable cost & Capacity factor & Lag & EM\\
				& (MW) & (\$/kW) & (\$/MWh) & & (year) & (kg/MWh)\\
				\midrule
				Hydro       & 9,213.0 & 1,200 &  3.4 & 0.45  & 5 & 0\\
				Wind        &   502.0 & 1,300 &  2.0 & 0.34  & 1 & 0\\
				Solar PV    &    10.0 &   900 &  0.0 & 0.196 & 1 & 0\\
				Geothermal  &     7.3 & 3,000 &  6.0 & 0.88  & 3 & 50\\
				Bioenergy   &    25.0 & 2,000 & 10.0 & 0.55  & 1 & 230\\
				\midrule
				\multicolumn{7}{c}{\textbf{Zimbabwe}}\\
				\midrule
				Technology & Initial capacity & CAPEX & Variable cost & Capacity factor & Lag & EM\\
				& (MW) & (\$/kW) & (\$/MWh) & & (year) & (kg/MWh)\\
				\midrule
				Hydro       & 1,119.6 & 1,691 &  2.0 & 0.50  & 5 & 0\\
				Coal        & 1,680.0 & 1,800 & 40.0 & 0.55  & 4 & 924\\
				Solar PV    &    70.1 &   900 &  0.0 & 0.203 & 1 & 0\\
				Gas         &     0.0 & 1,000 & 80.0 & 0.50  & 3 & 630\\
				Bioenergy   &    92.3 & 1,500 & 18.0 & 0.55  & 2 & 230\\
				\bottomrule
			\end{tabular*}
	\end{threeparttable}
\end{table}

The hydrological state process is derived from GloFAS version-5 river-discharge histories \citep{GloFASv5}.\footnote{GloFAS, the Global Flood Awareness System, is operated by the European Commission's Joint Research Centre and the European Centre for Medium-Range Weather Forecasts (ECMWF) under the Copernicus Emergency Management Service. Its historical product, the GloFAS-ERA5 reanalysis, is a continuous simulated record of river discharge, the volume of water passing a point on a river per unit of time. It gives a gap-free series for basins whose observed records are short.} These are the consolidated LISFLOOD\footnote{LISFLOOD is the hydrological and river-routing model underlying GloFAS. It converts gridded weather inputs, mainly precipitation, temperature, and evaporation, into runoff, and routes that runoff through the river network to give a discharge value at each cell.} reanalysis forced by ERA5\footnote{ERA5 is the fifth-generation global atmospheric reanalysis produced by ECMWF. A reanalysis blends historical observations with a numerical weather model to reconstruct a gap-free record of past conditions. Forced by ERA5 means that ERA5 supplies the weather inputs driving the hydrological simulation.}, providing average daily river discharge at 0.05\textdegree resolution over the 1980--2024 study period (45 complete years). For each country a gauge-consistent grid cell\footnote{A river gauge is a fixed station at which stream flow is measured. A gauge-consistent grid cell is the cell of the modeled river network corresponding to such a station, hence the simulated series represents flow at a recognized point on the river.} represents interannual basin conditions, the Blue Nile at El Diem (11.225\textdegree N, 34.975\textdegree E) for Ethiopia and the Zambezi at Victoria Falls ($-$17.925\textdegree, 25.825\textdegree) for Zimbabwe, are considered as basin-level proxies for the dominant hydropower systems. Annual mean flow is classified using the one-third and two-thirds empirical quantiles\footnote{The one-third and two-thirds empirical quantiles split the ranked historical years into three equally sized groups, such that the classification assigns one third of the record to each inflow condition by construction instead of imposing an external definition of a dry or wet year.} of the historical distribution into low-, normal-, and high-inflow conditions. This equal-frequency classification provides a representation of hydrological variability while maintaining a balanced historical representation across the three conditions. The condition multiplier $\mu_h$ is the mean normalized annual inflow within each condition, and the transition matrix $P^{H}$ is estimated from observed year-to-year transitions, and both are calibrated separately for each basin. The normalized multipliers are 0.84/1.01/1.17 for Ethiopia and 0.60/1.01/1.40 for Zimbabwe, scaled so that $\sum_h\pi_h\mu_h=1$. The initial condition $h_1$ is the low-inflow state in both countries, based on the 2024 observation immediately preceding the planning horizon. As an independent check, the GloFAS basin means exceed long-run estimates by about 12\% for Ethiopia \citep{Mekonnen2018} and about 5\% for Zimbabwe \citep{ZRA}. The state multipliers are normalized relative to each basin's long-run mean, and a purely multiplicative level difference largely cancels in the relative multipliers. Full multipliers, transition matrices, cutoff values, coordinates, the reanalysis-framework background, and the detailed validation arithmetic are provided in \ref{app_calibration}.

\subsection{Scenario design and evaluation}

The scenario design reflects the main planning uncertainties and policy choices facing the two power systems. For Ethiopia, the analysis examines changes in renewable investment costs and learning, faster electricity-access expansion, higher long-run demand, limits on new hydropower development, and sustained low-inflow conditions. The accelerated-access case raises demand to 50 TWh by 2030, while the hydrological stress case fixes each planning year in the low-inflow condition. Two Ethiopia cases also raise the planning constraints on hydropower, wind, and solar to 20, 15, and 15 GW, namely accelerated access and the high-demand case with expanded ceilings. The high-demand case retains the baseline capacity bounds, which separates the effect of higher demand from the effect of the bounds. The constrained-hydropower case, on the other hand, limits cumulative new hydropower construction to 3 GW. For Zimbabwe, the baseline places no restriction on carbon emissions and serves as the reference for changes in carbon policy, investment costs, demand, hydrological conditions, and the retirement of inherited coal capacity. Carbon taxes of \$30 and \$50/tCO$_2$ enter the planning objective function, while the associated tax payments are treated as domestic transfers in the resource cost accounting. The quantity-based cases include a glide-path cap that declines linearly from 8 to 3 MtCO$_2$ per year between 2025 and 2050 and a matched annual cap derived from the expected emissions trajectory under the reference carbon tax. A further case combines the \$50/tCO$_2$ tax with concessional solar financing. The financing assumption lowers the cost of capital from 12\% to 6\%, corresponding to an effective reduction of about 39\% in solar investment cost.

The learning-curve case for Ethiopia applies a multiplier to the baseline wind and solar investment cost trajectories that declines linearly from 1.00 in 2025 to 0.65 for wind and 0.55 for solar in 2050. It represents an exogenous alternative cost trajectory and not endogenous learning-by-doing, since no learning rate is estimated and cumulative deployment does not feed back into investment cost. Additional experiments examine the sensitivity of the expansion pathways to the discount rate, value of lost load, hydrological transitions, and the discretization of demand uncertainty. Separate uncertainty decomposition cases isolate the roles of demand and hydrological uncertainty, while the coal-retirement cases examine alternative retirement timing and the cost of continued operation of inherited coal capacity. Stochastic outcomes are evaluated using 5,000 independently generated out-of-sample paths, with 1,000 paths generated under each of five random seeds. Common paths are used when policies are compared directly to reduce sampling differences across paired evaluations. The matched-cap parameter is calibrated on a separate sample and subsequently evaluated on the same out-of-sample paths used for the other policy cases.

\section{Results}\label{sec_results}
This section presents the model results for both power systems. First, we compare the two countries' baseline installed capacities. Then we present detailed results for the Ethiopian and Zimbabwean GEP models, where the Zimbabwe case also includes emission-reduction policy comparisons. Finally, we present the adaptive policy produced by the SDDP model together with the cross-country comparison, which place the individual results for further implications.
 
\begin{figure}[htbp]
	\centering
	\begin{subfigure}[t]{0.45\textwidth}
		\centering
		\includegraphics[width=\textwidth]{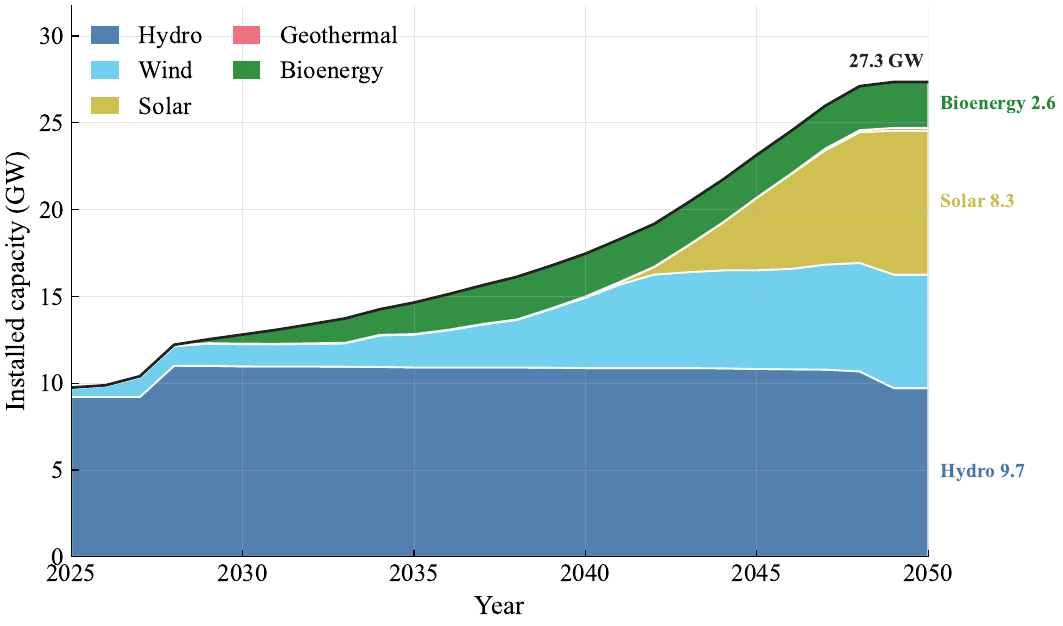}
	\end{subfigure}
	\hfill
	\begin{subfigure}[t]{0.45\textwidth}
		\centering
		\includegraphics[width=\textwidth]{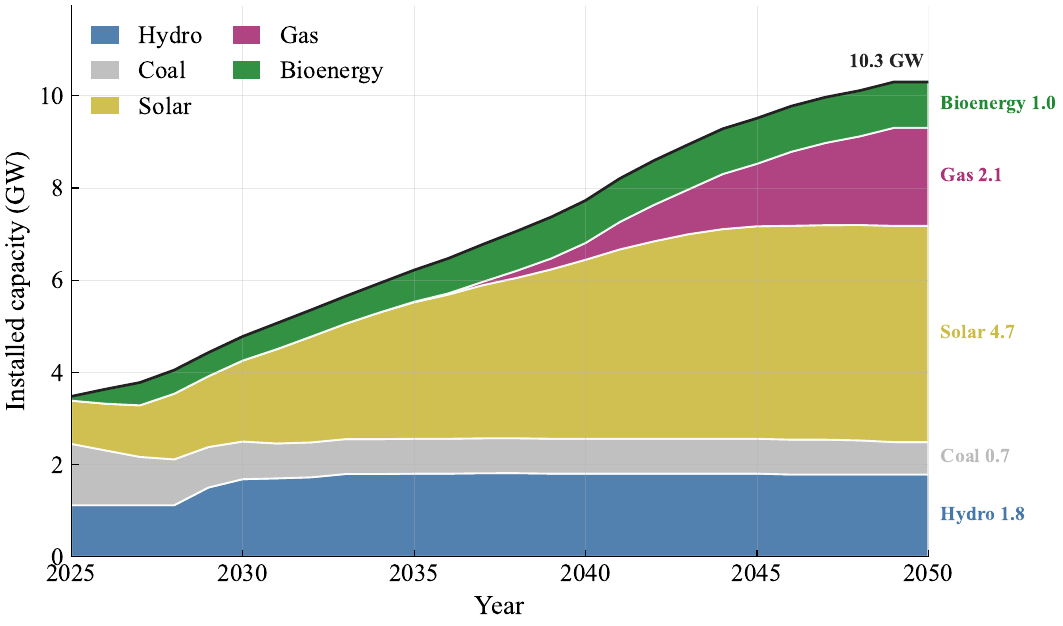}
	\end{subfigure}
	\caption{Baseline installed capacity by technology in Ethiopia (left) and Zimbabwe (right), 2025--2050.\label{fig_baseline_captraj}}
\end{figure}

Figure \ref{fig_baseline_captraj} compares the baseline capacity trajectories for the two systems. Ethiopia expands through a predominantly low-carbon mix of hydropower, wind, and solar, while Zimbabwe combines a larger solar expansion with retained coal and new gas capacity. Ethiopia's total capacity grows gradually before accelerating later in the horizon as wind and solar expansions increase, whereas Zimbabwe begins adding solar earlier while most scheduled coal retirement occurs near the start of the horizon.

\subsection{Ethiopia: expansion under demand and hydrological uncertainty}

\begin{figure}[htbp]
	\centering
	\includegraphics[width=0.98\textwidth]{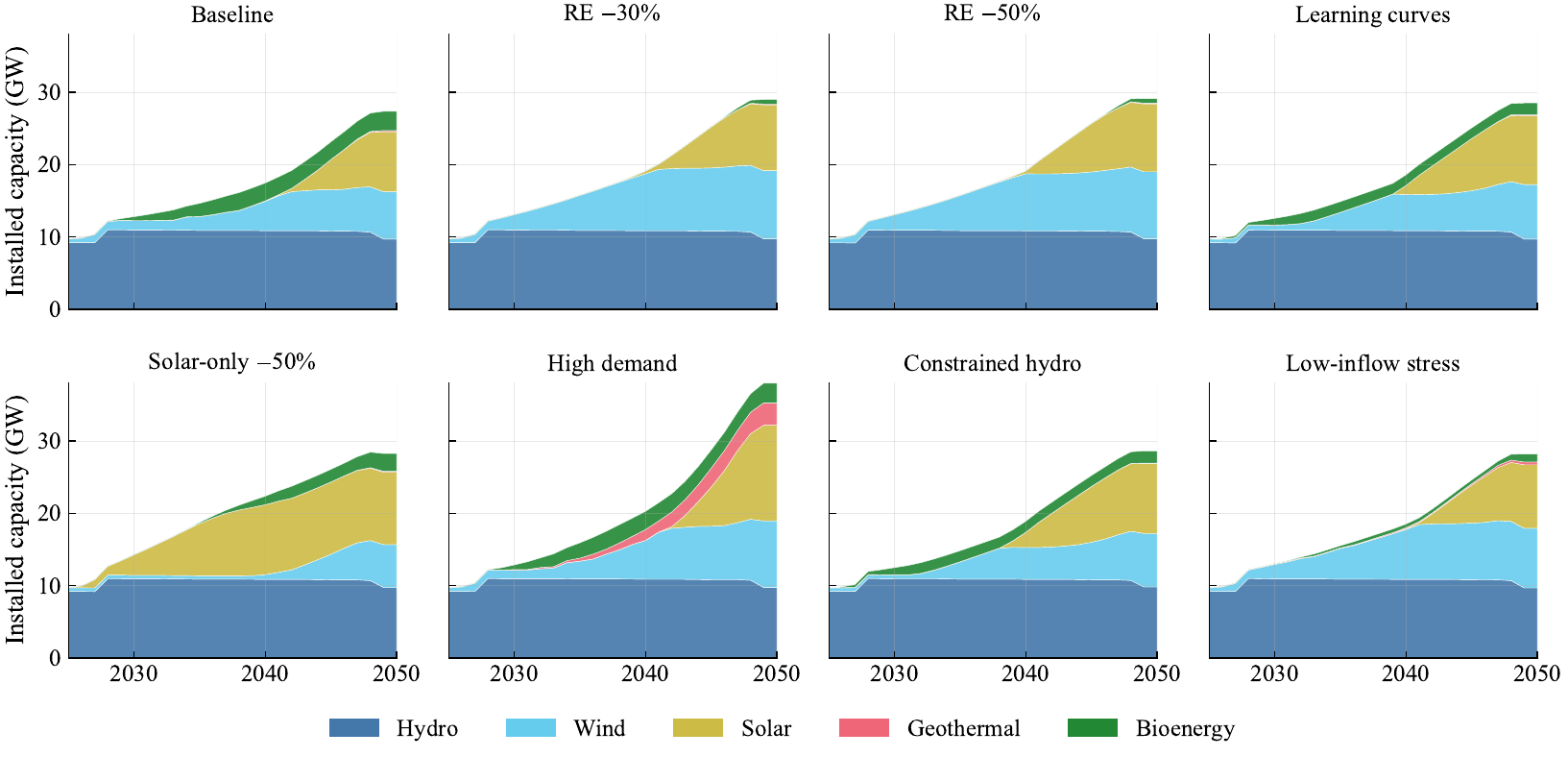}
	\caption{Installed capacity by technology across Ethiopia scenarios, 2025--2050.\label{fig_eth_capacity_traj}}
\end{figure}

\begin{figure}[htbp]
	\centering
	\includegraphics[width=0.98\textwidth]{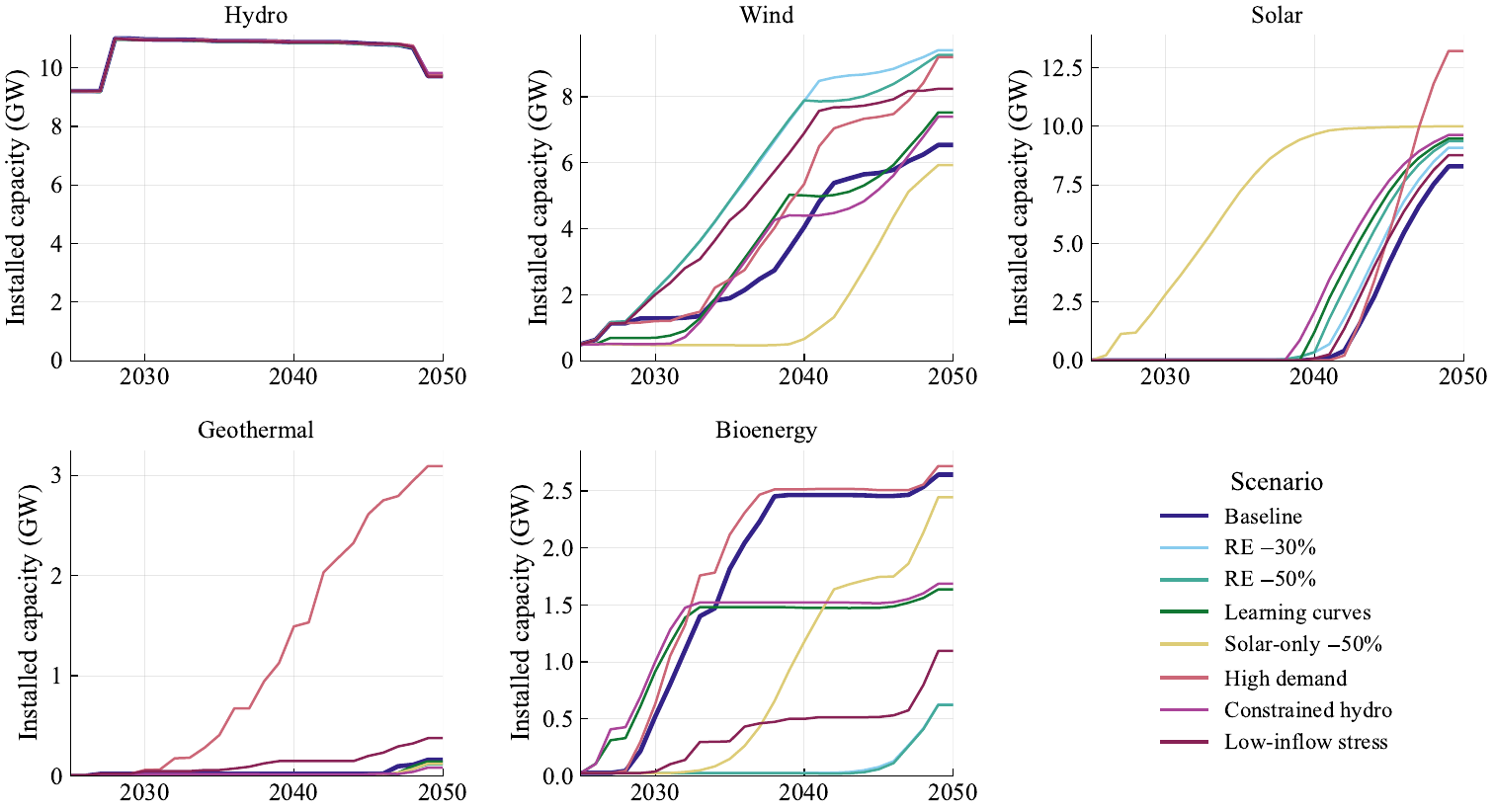}
	\caption{Technology-specific installed capacity trajectories across Ethiopia scenarios, 2025--2050.\label{fig_eth_build}}
\end{figure}

\begin{figure}[htbp]
	\centering
	\includegraphics[width=0.98\textwidth]{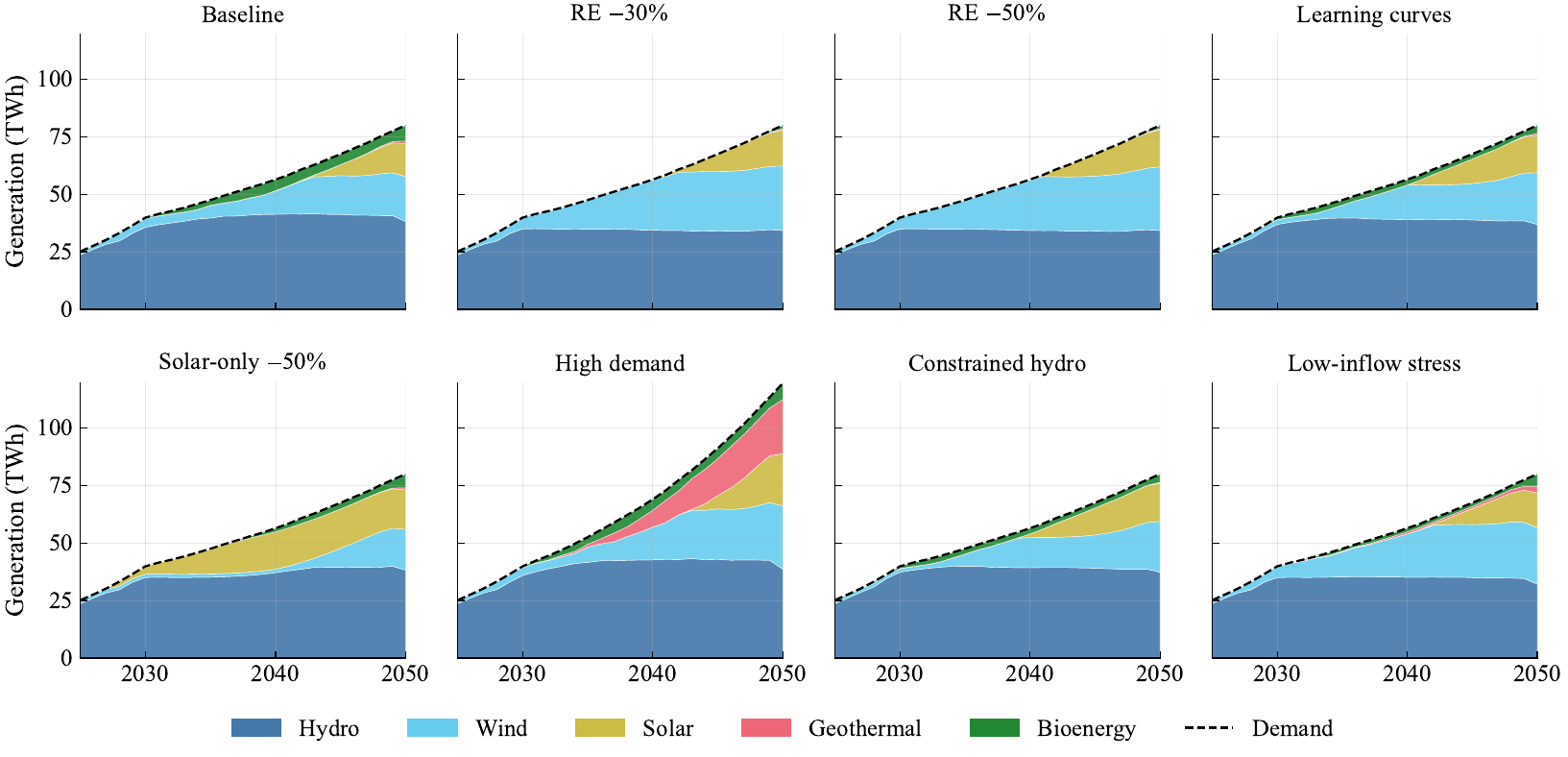}
	\caption{Generation by technology across Ethiopia scenarios, 2025--2050. The dashed line denotes realized demand.\label{fig_eth_genmix}}
\end{figure}

Figures \ref{fig_eth_capacity_traj} and \ref{fig_eth_build} show the Ethiopian baseline pathway, which combines the inherited hydropower fleet with a substantial expansion of wind and solar. Hydropower capacity rises to about 11 GW once Koysha commissions at the end of the 2020s and holds near that level until a scheduled retirement reduces it to about 9.7 GW in 2050, while wind and solar expand to about 6.5 and 8.3 GW by 2050, with smaller contributions from bioenergy and geothermal. The generation mix in Figure \ref{fig_eth_genmix} is led by hydropower at 38.2 TWh in 2050, close to 48\% of generation, followed by wind at 19.5 TWh and solar at 14.2 TWh. Figure \ref{fig_eth_build} shows that wind expands through much of the horizon, while most baseline solar expansions occur later. The renewable cost cases change both the pace and composition of this build, with the strongest shift occurring in wind.

Lower renewable investment costs shift the portfolio substantially toward wind. As Table \ref{tab_eth_snapshot} shows, reducing wind and solar investment cost by 30\% and 50\% raises wind capacity from 6.5 to about 9.4 GW and lowers gross resource cost from \$16.87 billion to \$14.68 and \$13.06 billion, while bioenergy capacity falls to about 0.6 GW and average annual emissions decline from 0.70 to 0.03 MtCO$_2$. The solar-only case provides a useful contrast. Solar reaches 10.0 GW in 2050, while wind remains below the baseline at 5.9 GW and bioenergy remains close to its baseline level at 2.4 GW. Broad renewable cost reductions produce a different substitution pattern, expanding both wind and solar while displacing most bioenergy. Restricting new hydropower shifts investment toward wind and solar, while geothermal capacity remains limited. Under high demand, gross resource cost rises to about \$24.75 billion as additional capacity is built to serve the larger load. The accelerated-access case also requires a large build, reaching 9.7 GW of wind and 12.1 GW of solar by 2050 at a gross resource cost of \$23.02 billion. The high-demand case differs in composition, with solar reaching 13.2 GW and geothermal 3.1 GW by 2050. The low-inflow stress case, which fixes every planning year to the low-inflow condition, lowers average annual emissions to 0.32 MtCO$_2$, with the reduced hydropower availability met by greater wind and geothermal generation and lower bioenergy generation.

\begin{table}[htbp]
	\centering
	\caption{Ethiopia scenario results for resource cost, 2050 capacity, average annual emissions, and expected unserved energy (EUE).\label{tab_eth_snapshot}}
	\footnotesize
		\begin{tabular*}{\textwidth}{@{\extracolsep{\fill}}lrrrrrrrr}
			\toprule
			Scenario & Cost & Hydro & Wind & Solar & Geothermal & Bioenergy & Average emission & EUE\\
			& (\$B) & (GW) & (GW) & (GW) & (GW) & (GW) & (Mt) & (TWh)\\
			\midrule
			Baseline & 16.87 & 9.7 & 6.5 & 8.3 & 0.2 & 2.6 & 0.70 & 0.05\\
			RE invest $-$30\% & 14.68 & 9.7 & 9.4 & 9.1 & 0.1 & 0.6 & 0.03 & 0.05\\
			RE invest $-$50\% & 13.06 & 9.7 & 9.3 & 9.4 & 0.1 & 0.6 & 0.03 & 0.05\\
			Learning curves & 15.60 & 9.7 & 7.5 & 9.5 & 0.1 & 1.6 & 0.42 & 0.05\\
			Solar-only $-$50\% & 14.89 & 9.8 & 5.9 & 10.0 & 0.1 & 2.4 & 0.30 & 0.05\\
			Constrained hydro & 15.55 & 9.8 & 7.4 & 9.6 & 0.1 & 1.7 & 0.44 & 0.06\\
			High demand $+$50\% & 24.75 & 9.7 & 9.2 & 13.2 & 3.1 & 2.7 & 1.10 & 0.06\\
			Low-inflow stress & 17.15 & 9.7 & 8.2 & 8.8 & 0.4 & 1.1 & 0.32 & 0.05\\
			Accelerated access & 23.02 & 9.7 & 9.7 & 12.1 & 0.2 & 2.7 & 0.81 & 0.04\\
			\bottomrule
		\end{tabular*}
\end{table}

\begin{figure}[htbp]
	\centering
	\begin{subfigure}[t]{0.49\textwidth}
		\centering
		\includegraphics[width=\textwidth]{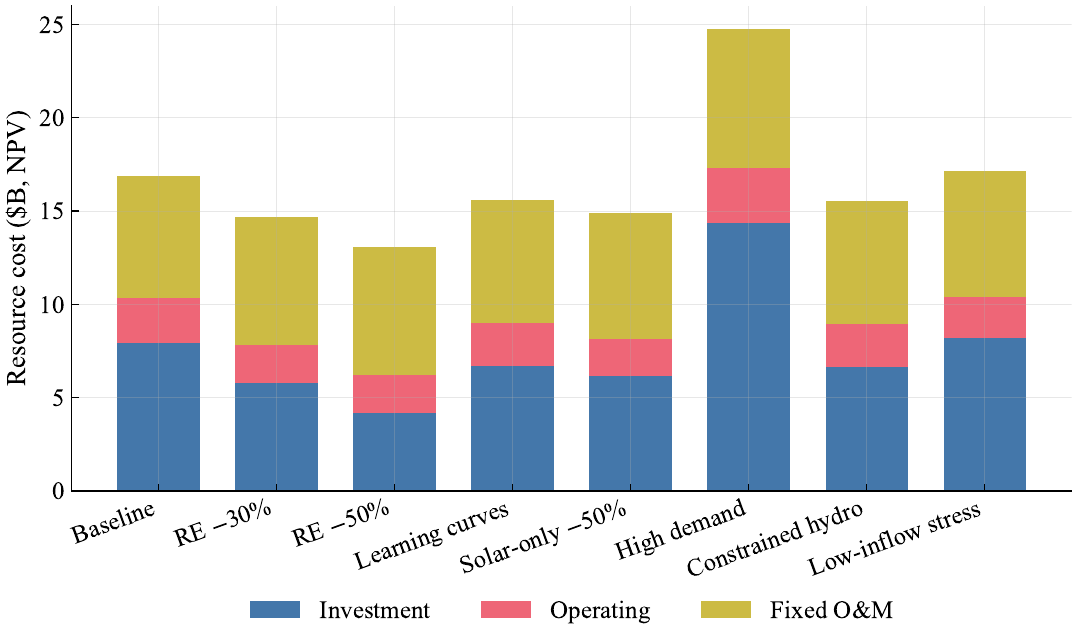}
		\caption{Ethiopia}
	\end{subfigure}
	\hfill
	\begin{subfigure}[t]{0.49\textwidth}
		\centering
		\includegraphics[width=\textwidth]{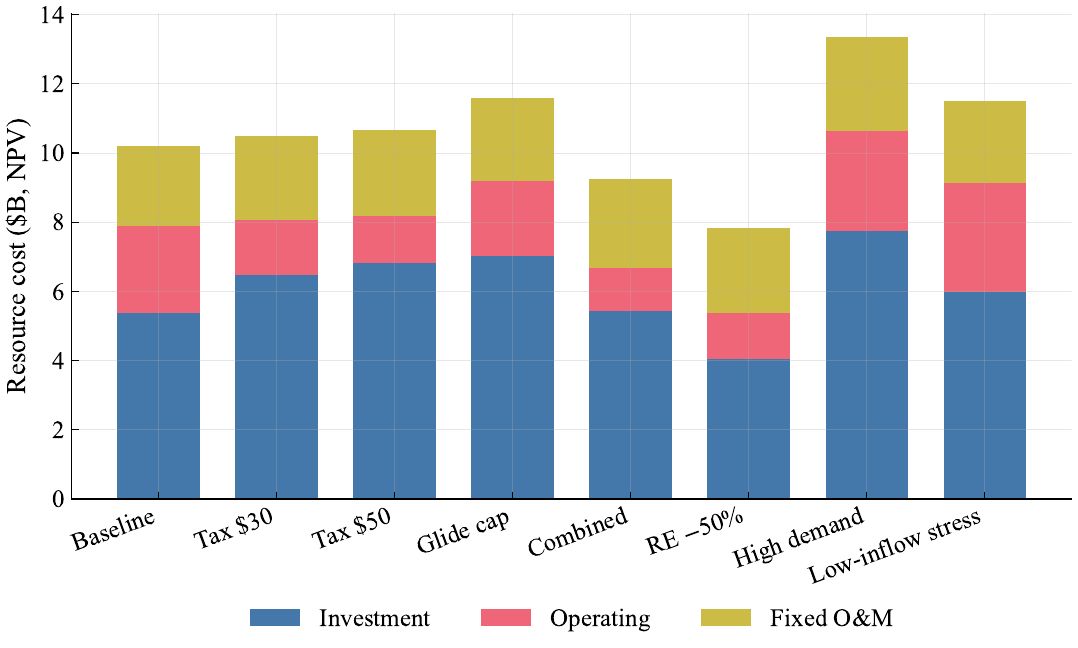}
		\caption{Zimbabwe}
	\end{subfigure}
	\caption{Present value resource cost decomposition by scenario for Ethiopia and Zimbabwe.\label{fig_cost_decomp}}
\end{figure}

Baseline gross resource cost is \$16.87 billion, and net resource cost after salvage value is \$15.59 billion. The present value resource cost decomposition in Figure \ref{fig_cost_decomp} ranges from about \$13 to \$25 billion across the Ethiopia scenarios, with the largest value in the high-demand case. Investment and fixed O\&M account for most of the resource cost across these scenarios, while variable operating cost remains comparatively small. The high-demand case increases both investment expenditure and the fixed cost of the larger fleet. Across the scenarios examined, expected unserved energy remains small relative to total demand, at 0.05 TWh in the baseline and no more than about 0.06 TWh.

\subsection{Zimbabwe: baseline pathway and coal retirement}

\begin{figure}[htbp]
	\centering
	\includegraphics[width=0.98\textwidth]{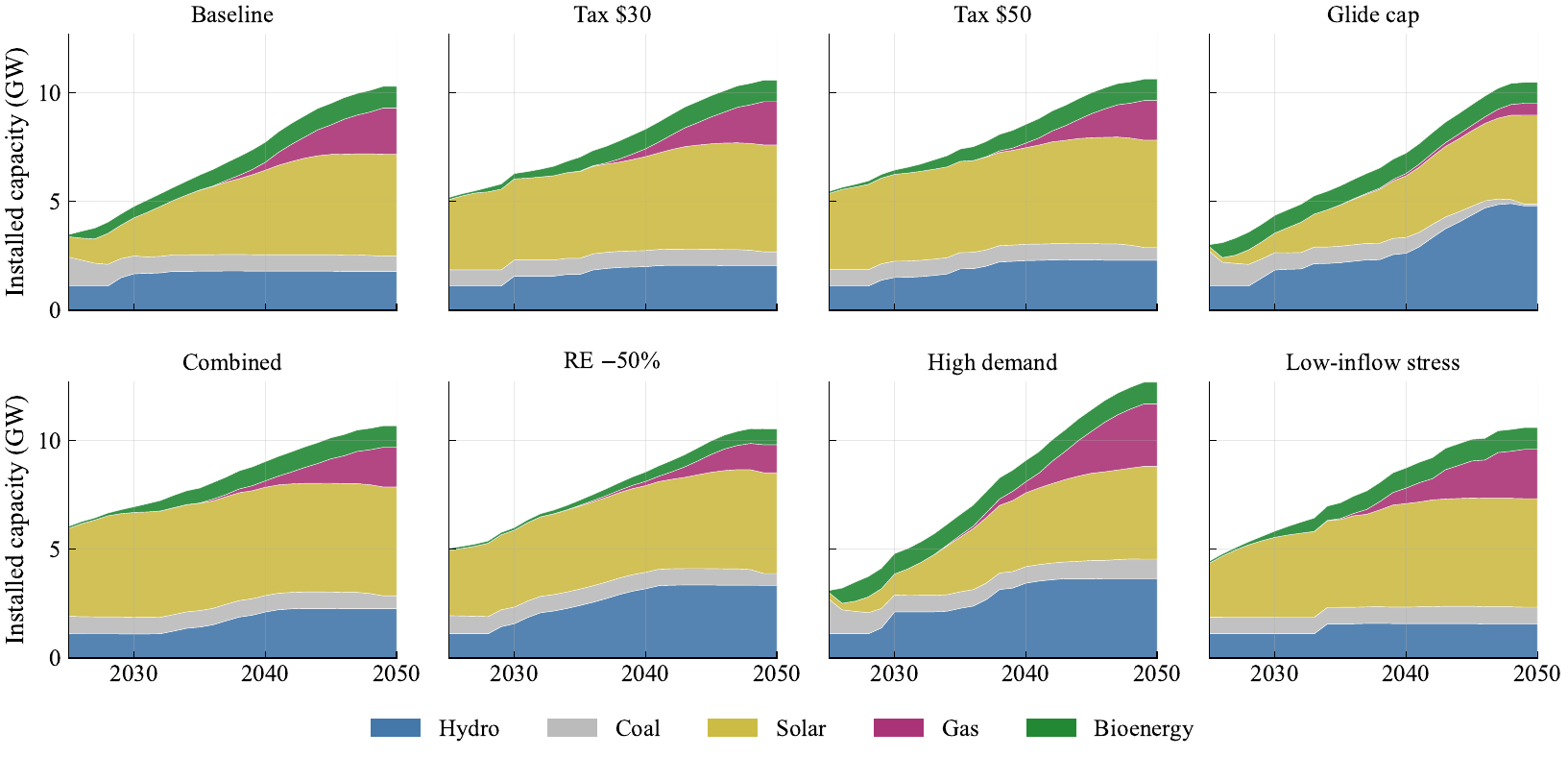}
	\caption{Installed capacity by technology across Zimbabwe scenarios, 2025--2050.\label{fig_zwe_genmix_scenario}}
\end{figure}

\begin{figure}[htbp]
	\centering
	\includegraphics[width=0.98\textwidth]{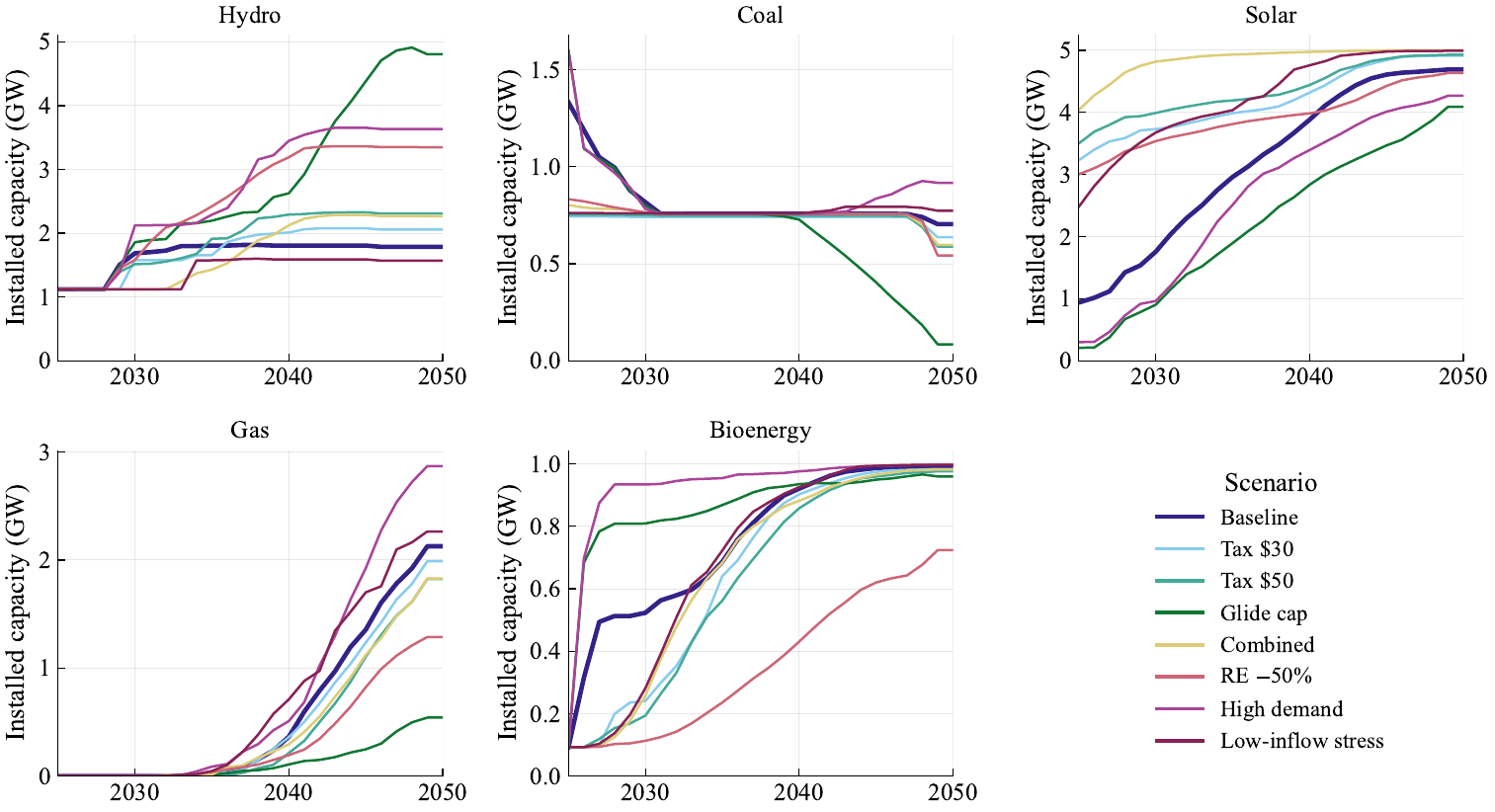}
	\caption{Technology-specific installed-capacity trajectories across Zimbabwe scenarios, 2025--2050.\label{fig_zwe_build}}
\end{figure}

Figures \ref{fig_zwe_genmix_scenario} and \ref{fig_zwe_build} show the Zimbabwe baseline pathway. Scheduled retirement reduces the inherited Hwange coal fleet from 1.68 GW in 2025 to about 0.70 GW by 2050, while the model adds about 2.1 GW of gas and expands solar to about 4.7 GW alongside a hydropower base of about 1.8 GW. Solar and hydropower are the two largest sources of generation in 2050, at 8.3 and 7.9 TWh, followed by bioenergy, coal, and gas. Figure \ref{fig_zwe_build} clarifies the sequencing of the baseline pathway. Coal falls rapidly toward its retirement floor by the early 2030s, solar expands from the early years, and most gas additions occur later in the horizon.

The baseline applies no carbon tax and no emissions cap. Gross resource cost is \$10.21 billion, and net resource cost after salvage value is \$9.69 billion. Average annual emissions are 2.80 MtCO$_2$ and reach 5.31 MtCO$_2$ in 2050 as demand growth raises generation from the remaining thermal fleet. Expected unserved energy is 0.04 TWh, small relative to total demand.

\subsection{Zimbabwe: carbon policy}

\begin{figure}[H]
	\centering
	\includegraphics[width=0.7\textwidth]{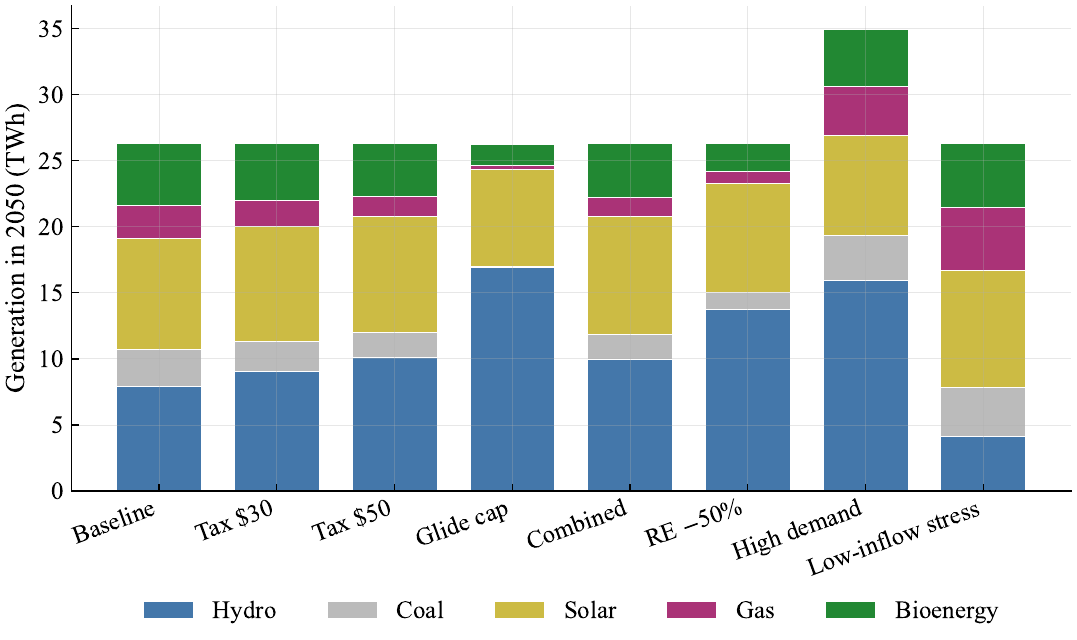}
	\caption{Generation by technology across Zimbabwe scenarios in 2050.\label{fig_zwe_genmix2050}}
\end{figure}

Table \ref{tab_zwe_policy} summarizes the Zimbabwe policy outcomes. At \$30/tCO$_2$, the carbon tax raises gross resource cost by 2.7\% and reduces average annual emissions from 2.80 to 1.78 MtCO$_2$. At \$50/tCO$_2$, the premium rises to 4.5\% and average emissions fall to 1.47 MtCO$_2$. Figure \ref{fig_zwe_genmix2050} shows that both taxes reduce coal and gas generation and raise solar generation in 2050. Figures \ref{fig_zwe_genmix_scenario} and \ref{fig_zwe_build} show that the policy effect also appears in investment timing, with solar reaching high capacity earlier under the tax cases and late gas additions remaining below the baseline. The glide-path cap, which tightens the annual ceiling from 8 to 3 MtCO$_2$ over the horizon, reaches the lowest final year emissions at 0.71 MtCO$_2$ by displacing almost all coal and expanding hydropower, at a higher gross resource cost of \$11.58 billion and higher expected unserved energy of 0.17 TWh. The high-demand and low-inflow cases instead raise emissions. Their 2050 emissions rise to 6.42 and 7.53 MtCO$_2$, respectively. Figure \ref{fig_zwe_genmix2050} shows greater thermal generation when additional demand or reduced hydropower availability has to be covered.

\begin{table}[htbp]
	\centering
	\caption{Zimbabwe policy outcomes and gross resource cost premiums relative to the baseline.\label{tab_zwe_policy}}
	\footnotesize
		\begin{tabular*}{\textwidth}{@{\extracolsep{\fill}}lrrrrrrr}
			\toprule
			Scenario & Gross resource & Net resource & Cash & Average emission & 2050 emission & EUE & Premium\\
			& cost (\$B) & cost (\$B) & (\$B) & (Mt) & (Mt) & (TWh) & (\%)\\
			\midrule
			Baseline (no policy) & 10.21 & 9.69 & 10.21 & 2.80 & 5.31 & 0.04 & +0.0\\
			Carbon tax \$30/t & 10.49 & 9.93 & 11.26 & 1.78 & 4.28 & 0.04 & +2.7\\
			Carbon tax \$50/t & 10.66 & 10.09 & 11.74 & 1.47 & 3.63 & 0.05 & +4.5\\
			Glide cap 8$\to$3 Mt & 11.58 & 10.77 & 11.58 & 1.94 & 0.71 & 0.17 & +13.4\\
			Tax \$50 + Solar fin. & 9.24 & 8.73 & 10.19 & 1.33 & 3.62 & 0.04 & -9.5\\
			RE invest $-$50\% & 7.83 & 7.43 & 7.83 & 1.34 & 2.22 & 0.04 & -23.3\\
			High demand & 13.36 & 12.50 & 13.36 & 2.99 & 6.42 & 0.10 & +30.8\\
			Low-inflow stress & 11.51 & 10.99 & 11.51 & 3.84 & 7.53 & 0.07 & +12.7\\
			Matched cap (\$30) & 10.59 & 10.03 & 10.59 & 1.82 & 4.14 & 0.06 & +3.7\\
			Matched cap (\$50) & 10.89 & 10.30 & 10.89 & 1.45 & 3.42 & 0.07 & +6.7\\
			\bottomrule
		\end{tabular*}
\end{table}

At approximately matched expected emissions, the annual cap has a higher net resource cost than the corresponding carbon tax. The difference in net resource cost is about \$0.10 billion at the \$30 reference and \$0.21 billion at the \$50 reference, and expected unserved energy under the caps is 0.06 and 0.07 TWh, higher than the 0.04 and 0.05 TWh under the corresponding taxes. Out-of-sample emissions at the matched points differ by less than 2\% between the two instruments. This comparison is specific to the modeled Zimbabwe system and does not imply a general ranking of price and quantity instruments. On the presented mean annual emissions path, none of the three cap cases reaches its annual ceiling in any year. The glide-path cap comes closest in 2025, when mean annual emissions of 7.09 MtCO$_2$ stand against an 8 MtCO$_2$ ceiling, a utilization of 88.6\%. Utilization falls to 33.4\% in 2030 and stays below 30\% thereafter, reaching 23.6\% in 2050. The matched annual caps sit further from their limits on the mean path, with maximum utilization of 58.3\% at the \$30 reference and 49.0\% at the \$50 reference, both in 2026. The constraints apply to realized annual emissions, where slack on the mean path does not establish slack in every stochastic realization, and the archived production summaries do not retain pathwise binding frequencies.

\begin{figure}[htbp]
	\centering
	\includegraphics[width=0.62\textwidth]{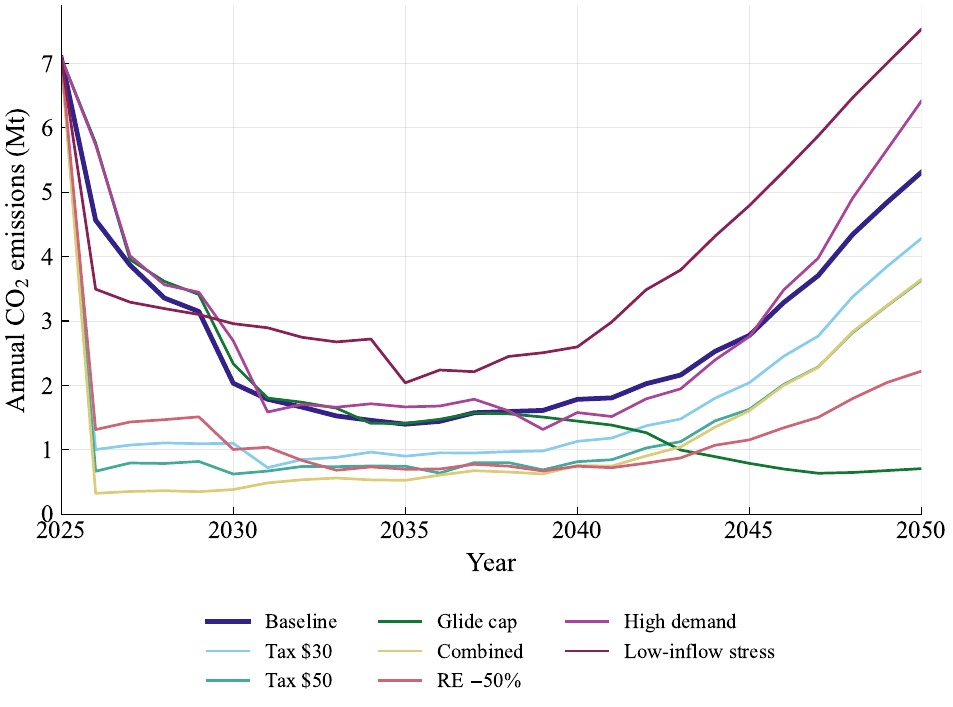}
	\caption{Annual CO$_2$ emissions across Zimbabwe policy scenarios.\label{fig_zwe_emissions_compare}}
\end{figure}

Combining the \$50/tCO$_2$ tax with concessional solar financing, an effective reduction of about 39\% in the solar capital cost, lowers gross resource cost to \$9.24 billion, below the baseline, while reducing average emissions to 1.33 MtCO$_2$. A 50\% reduction in renewable investment cost lowers gross resource cost to \$7.83 billion. Lower solar and renewable investment costs offset the cost increase associated with carbon pricing while also reducing emissions. The annual emissions trajectories in Figure \ref{fig_zwe_emissions_compare} share a steep decline as the scheduled Hwange retirements occur over 2026--2031, after which the policy paths separate. Baseline and high-demand emissions rise strongly with demand, the tax paths increase more gradually, and the glide cap continues to suppress emissions toward the end of the horizon.

\subsection{Value of adaptive planning and shortage exposure}\label{sec_adaptivity}

Table \ref{tab_vss} compares the multistage policy with the deterministic and perfect information benchmarks, with all three plans scored on the same 200 common out-of-sample paths. The deterministic plan fixes investment before any uncertainty is realized and cannot adjust to later demand or/and hydrological outcomes. Its fixed schedule leaves large expected unserved energy and a much higher total expected cost. The value of adaptive planning, measured as the reduction in total expected cost relative to the deterministic plan is about \$362 billion for Ethiopia and \$80 billion for Zimbabwe, with 95\% confidence interval half-widths of about \$27 and \$7 billion. The value of perfect information is much smaller, at about \$1.5 billion for Ethiopia and \$1.0 billion for Zimbabwe. The magnitude of these differences is driven primarily by the shortage cost of the deterministic plan at the assumed value of lost load, and they are not a direct forecast of realized gains from adopting the method. These are differences in total expected cost. The deterministic plan attains a lower gross resource cost in both countries by under-building, and the comparison is dominated by the resulting shortage valuation.

\begin{table}[htbp]
	\centering
	\caption{Out-of-sample benchmarks for the value of adaptive planning.\label{tab_vss}}
	\footnotesize
		\begin{tabular*}{\textwidth}{@{\extracolsep{\fill}}lrrrr}
			\toprule
			Plan & Gross resource & Expected unserved & Salvage value & Total expected\\
			& cost (\$B, NPV) & energy (TWh) & credit (\$B) & cost (\$B)\\
			\midrule
			\multicolumn{5}{@{}l}{\textit{Ethiopia}}\\
			\quad Perfect information & 15.29 & 0.00 & 1.13 & 14.16\\
			\quad Stochastic (SDDP) & 16.83 & 0.04 & 1.26 & 15.70\\
			\quad Deterministic (EV) & 13.17 & 78.09 & 0.97 & 377.86\\
			\midrule
			\multicolumn{5}{@{}l}{\textit{Zimbabwe}}\\
			\quad Perfect information & 9.20 & 0.00 & 0.43 & 8.78\\
			\quad Stochastic (SDDP) & 10.24 & 0.02 & 0.52 & 9.81\\
			\quad Deterministic (EV) & 9.47 & 15.90 & 0.53 & 90.08\\
			\bottomrule
		\end{tabular*}
\end{table}

\begin{figure}[htbp]
	\centering
	\includegraphics[width=0.90\textwidth]{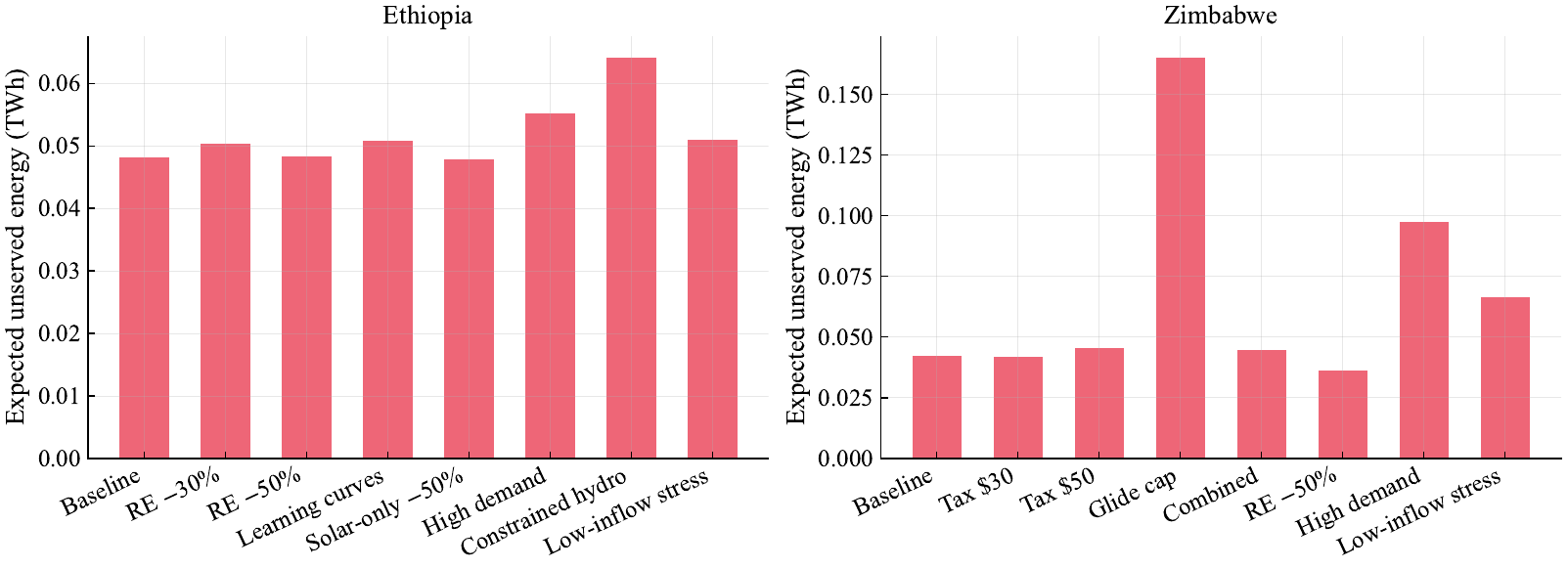}
	\caption{Expected unserved energy across the main Ethiopia and Zimbabwe scenarios.\label{fig_adequacy}}
\end{figure}

The physical effect of adaptive investment is clear in expected unserved energy. Relative to the deterministic plan, adaptive investment lowers expected unserved energy from 78.1 TWh to 0.04 TWh in Ethiopia and from 15.9 TWh to 0.02 TWh in Zimbabwe over the planning horizon. Figure \ref{fig_adequacy} shows that expected unserved energy under the adaptive plan remains small across the main scenarios. The main exception is the Zimbabwe glide cap, where expected unserved energy (EUE) rises to 0.17 TWh, followed by the high-demand case at 0.10 TWh. The remaining main Zimbabwe cases lie between 0.04 and 0.07 TWh. On the other hand, the Ethiopia cases remain tightly clustered between 0.04 and 0.06 TWh. Table \ref{tab_vollvss} revalues the same plans at alternative values of lost load without re-optimizing investment. The avoided expected unserved energy is fixed at 78.05 TWh for Ethiopia and 15.88 TWh for Zimbabwe, and only its valuation moves, spanning about \$179 to \$728 billion and \$40 to \$161 billion, respectively. These magnitudes are valuations of avoided shortage on the total expected cost basis and are not comparable to the gross resource cost of about \$17 billion. The adaptive plan spends about \$3.7 billion more on capacity and operation than the deterministic plan, and its value is the much larger shortage cost that this additional spending avoids.

Table \ref{tab_uncdecomp} separates the two sources of uncertainty and shows that hydrological uncertainty has the larger effect on gross resource cost in both systems. With no uncertainty represented, gross resource cost is \$13.39 billion for Ethiopia and \$8.64 billion for Zimbabwe. Representing hydrological uncertainty alone raises it to \$15.88 and \$9.42 billion, and representing demand uncertainty alone raises it to \$14.29 and \$9.16 billion. In both countries the hydrology-only case moves gross resource cost closer to the joint-uncertainty outcome than the demand-only case, with the difference particularly higher in Ethiopia.

\subsection{Cross-country comparison and synthesis}

\begin{figure}[htbp]
	\centering
	\includegraphics[width=0.92\textwidth]{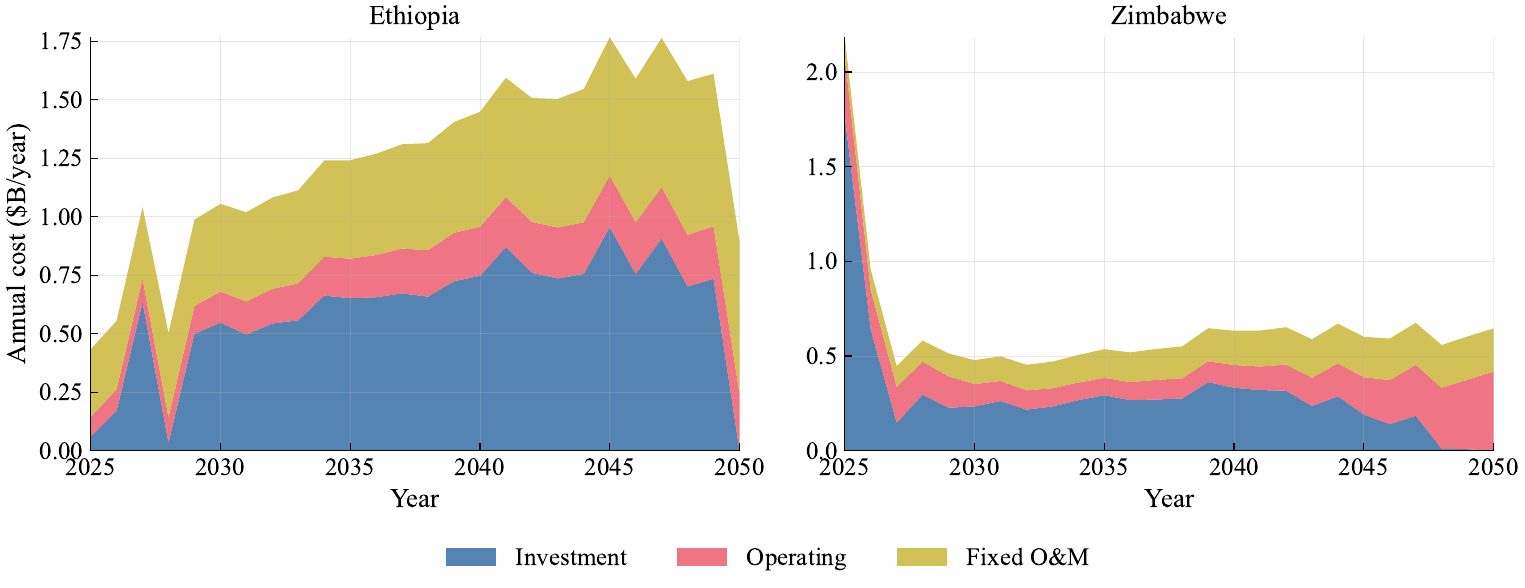}
	\caption{Baseline annual system cost by component for Ethiopia and Zimbabwe.}
	\label{fig_eth_zwe_cost_annual}
\end{figure}

Table \ref{tab_cross_country} compares the baseline outcomes for the two systems on a common modeling basis. Ethiopia reaches about 27.3 GW of installed capacity and 79.9 TWh of generation by 2050 as demand rises from 25.18 TWh in 2025 to 79.6 TWh. Zimbabwe reaches about 10.3 GW of installed capacity and 26.3 TWh of generation by 2050, from a system supply requirement of 11.082 TWh in 2025. Ethiopia's expansion is predominantly low-carbon, combining a large hydropower with substantial wind and solar, while Zimbabwe retains inherited coal, adds gas, and expands solar. The annual cost profiles in Figure \ref{fig_eth_zwe_cost_annual} show Ethiopia's expenditure rising steadily with demand, while Zimbabwe commits more of its investment earlier and then follows a lower annual profile. The composition also differs. Ethiopia carries recurrent investment and rising fixed O\&M as the generation fleet expands, whereas Zimbabwe is more front-loaded and operating cost forms a larger share of expenditure later in the horizon.

\begin{table}[H]
	\centering
	\caption{Baseline comparison of the Ethiopia and Zimbabwe power systems.\label{tab_cross_country}}
	\small
	\begin{tabular}{lrr}
		\toprule
		Metric & Ethiopia & Zimbabwe\\
		\midrule
		Planning horizon & 2025--2050 & 2025--2050\\
		Technologies & 5 & 5\\
		2050 capacity (GW) & 27.3 & 10.3\\
		2050 generation (TWh) & 79.9 & 26.3\\
		Gross resource cost (\$B) & 16.87 & 10.21\\
		Average annual emissions (Mt) & 0.70 & 2.80\\
		Expected unserved energy (TWh) & 0.05 & 0.04\\
		\bottomrule
	\end{tabular}
\end{table}

The one-at-a-time sensitivity analyses in Table \ref{tab_sensitivity} show that the social discount rate has the largest effect on present value resource cost. As the rate moves from 3\% to 8\%, resource cost ranges from \$21.36 to \$12.30 billion for Ethiopia and from \$12.15 to \$8.00 billion for Zimbabwe. This implies the weight placed on capital-intensive investment over a long horizon. Changes in the value of lost load and in the low-inflow hydrology multiplier produce smaller variations in resource cost and expected unserved energy. The VoLL sensitivity is expressed more clearly in shortage exposure than in gross resource cost. Lowering VoLL to \$5,000/MWh raises EUE to 0.08 TWh in Ethiopia and 0.10 TWh in Zimbabwe, while \$20,000/MWh reduces it to 0.02 TWh in both systems. The low-inflow multiplier has a much larger effect on resource cost in Ethiopia than in Zimbabwe, consistent with Ethiopia's greater hydropower exposure.

\begin{table}[htbp]
	\centering
	\caption{One-at-a-time sensitivity of baseline outcomes to selected model parameters.\label{tab_sensitivity}}
	\small
	\begin{tabular}{lrrrr}
		\toprule
		Sensitivity & \multicolumn{2}{c}{Ethiopia} & \multicolumn{2}{c}{Zimbabwe}\\
		\cmidrule(lr){2-3}\cmidrule(lr){4-5} & Cost (\$B) & EUE (TWh) & Cost (\$B) & EUE (TWh)\\
		\midrule
		Baseline (VoLL \$10,000/MWh) & 16.87 & 0.05 & 10.21 & 0.04\\
		\midrule
		Discount rate 3\% & 21.36 & 0.03 & 12.15 & 0.03\\
		Discount rate 8\% & 12.30 & 0.07 & 8.00 & 0.06\\
		VoLL \$5,000/MWh & 16.79 & 0.08 & 10.10 & 0.10\\
		VoLL \$20,000/MWh & 16.93 & 0.02 & 10.30 & 0.02\\
		Low-inflow multiplier $-$15\% & 19.49 & 0.05 & 10.50 & 0.05\\
		Low-inflow multiplier $+$15\% & 14.72 & 0.03 & 9.95 & 0.04\\
		\bottomrule
	\end{tabular}
\end{table}

\section{Discussion}\label{sec_discussion}

The central finding is that adaptive investment has substantial value under demand and hydrological uncertainty. A deterministic expected value plan commits the full investment schedule before uncertainty is realized, leaving no opportunity to revise future capacity commitments as demand and hydrological conditions evolve. This fixed schedule creates substantial exposure to unserved energy in both countries. Adaptive planning instead allows later investment decisions to respond to the observed system state, which reduces shortage exposure. Construction lead times limit the speed of this response, since newly observed conditions cannot immediately be converted into commissioned capacity. They do not eliminate the value of adaptation, as subsequent investment commitments can still be revised as the system evolves.

The benchmark results distinguish the value of adaptation from the value of foresight. In both applications, the value of adaptive planning is larger than the value of perfect information. In this regard, most of the measured benefit comes from retaining the ability to adjust future investment decisions as uncertainty unfolds, with full knowledge of the entire future path providing a smaller additional gain once this flexibility is available. This ordering is specific to the two applications and to the benchmark definitions used in the analysis. The physical and monetary measures also require different interpretations. The physical effect is reflected in the large reduction in expected unserved energy under the adaptive policy. The monetary value is measured through the difference in total expected cost and is driven mainly by the shortage cost incurred by the deterministic plan at the adopted value of lost load. Therefore, it should be interpreted as a stochastic programming benchmark under the assumed shortage valuation.

The two applications also illustrate how differences in the initial generation system shape the investment response. In the one-at-a-time decomposition, hydrological uncertainty has the larger effect on gross resource cost in both countries. For Ethiopia, this result is closely linked to the large role of hydropower, whose annual contribution varies directly with water availability. The system starts from a large inherited hydropower base and becomes increasingly diversified as demand grows. Wind and solar account for substantial additions even though hydropower remains the largest single source of generation in 2050. The portfolio is also responsive to renewable investment costs. Lower wind and solar costs shift expansion significantly toward wind, showing that the economics of renewable investment influence the least-cost pathway alongside hydropower development. The low-inflow stress case provides a related example of portfolio adjustment. Reduced hydropower availability is accompanied by greater wind and geothermal generation and substantially lower bioenergy generation. Given the positive net emissions coefficient assigned to bioenergy, this substitution lowers average emissions in the modeled case. The result is specific to the available technologies, investment costs, and resource limits and should not be interpreted as a general relationship between lower hydrological inflow and emissions.

Zimbabwe follows a different pathway from its initial hydro-thermal system. Hydrological uncertainty remains economically relevant despite the more diversified generation portfolio. Scheduled retirement removes a substantial share of the inherited Hwange coal fleet during the early part of the horizon, with some coal retained in the baseline alongside new gas investment and a large expansion of solar. The coal retirement decomposition clarifies the role of the inherited fleet in this pathway. Introducing a recurring continued-operation cost for coal leads the model to retire nearly all remaining coal capacity, expand gas further, and reduce emissions at a higher resource cost. The long-term pathway is sensitive to the assumed economic life and continued operating cost of inherited coal assets. Retention of thermal capacity in the baseline should consequently be understood as an outcome of the adopted cost and retirement assumptions.

Furthermore, the Zimbabwe carbon policy analysis compares price and quantity instruments on a common emissions basis. Carbon-tax payments are excluded from resource cost and recorded separately as transfers within the domestic account. This accounting treatment allows the tax and cap cases to be compared in terms of the real resources used by the power system without implying that tax payments lack fiscal and distributional consequences. Calibrating the instruments to approximately matched expected emissions also limits differences arising simply from unequal policy stringency. Under this comparison, the annual cap has a higher net resource cost and expected unserved energy than the corresponding carbon tax. The annual ceiling can become more restrictive in low-inflow years, since emissions cannot adjust beyond the prescribed limit. Whereas emissions under the tax respond endogenously to system conditions at the given carbon price. The differences remain modest and apply to the modeled Zimbabwe system, its technology options, and the policy cases considered. They should not be established as a general ranking between price and quantity instruments.

A separate result concerns the interaction between carbon policy and the cost of replacement capacity. Carbon pricing changes the incentive to reduce emitting generation, while investment costs and financing conditions determine the resource cost of the capacity response. In the modeled Zimbabwe system, combining the higher carbon tax with concessional solar financing reduces gross resource cost below the baseline. Lower renewable investment costs produce an additional reduction in resource cost. These results indicate that the economic consequences of carbon policy depend jointly on policy stringency and the investment environment. This distinction is particularly relevant in a capital-intensive expansion setting, where the cost of financing new low-carbon capacity can significantly affect the resource cost associated with emissions reduction.

\subsection{Insights and policy implications}

For Ethiopia, hydropower development and renewable diversification form part of the same long-term planning problem. The large inherited hydropower base exposes future supply to changes in water availability, which increases the value of preserving flexibility in later investment decisions. Similarly, the sensitivity to wind and solar investment costs shows that diversification is not determined by hydrological conditions alone. Lower renewable investment costs also alter the least-cost portfolio and increase the role of wind and solar as demand expands. The planning implication is not a choice between hydropower and variable renewables, but the need to consider their joint evolution under uncertain hydrology, growing demand, and changing investment costs. The low-inflow experiment reinforces this point by showing that the composition of the substitution away from hydropower also affects the emissions pathway.

For Zimbabwe, three considerations shape the long-term investment pathway. The first is the scheduled retirement and continued economic life of inherited coal capacity. The second is the design and stringency of carbon policy. The third is the cost and financing of the capacity that replaces emitting generation. These influences are related, although they operate through different channels. The baseline retains some coal while expanding gas and solar, whereas imposing a recurring continued-operation cost shifts the solution toward near-complete coal retirement and greater gas investment. Carbon pricing lowers emissions, and concessional solar financing reduces both resource cost and emissions relative to the baseline in the modeled case. The results show that coal retirement assumptions, carbon policy, and financing conditions need to be interpreted jointly when assessing the modeled transition. They do not prescribe a particular retirement schedule or imply that the resulting gas investment constitutes a policy recommendation.

Several model features define the scope of these conclusions. The annual formulation is intended for long-term investment planning and does not represent hourly renewable variability, unit commitment, ramping, within-year reservoir operation, and storage cycling. Thus, expected unserved energy should be interpreted as an annual energy-shortage measure. Hydrology is represented through annual basin-level Markov conditions that capture interannual variation in hydropower availability. Reservoir-specific operation, cascade coordination, seasonal water allocation, and competing water uses remain outside the formulation, which we leave as directions for future work.

The spatial and investment representations impose additional boundaries. The national single-node formulation omits internal transmission constraints and endogenous cross-border electricity trade, with Zimbabwean imports embedded in the calibrated system-supply requirement than modeled as an investment and dispatch option. Regional coordination could alter domestic investment and shortage outcomes when hydrological conditions differ across interconnected systems \citep{Spalding2017}. Future technology costs are deterministic inputs, and continuous investment decisions approximate large indivisible hydropower and thermal projects instead of representing discrete project choices \citep{Zou2019}. The emissions results also depend partly on the positive net coefficient assigned to bioenergy, particularly in the Ethiopian low-inflow case. Finally, project delivery and regulatory and governance conditions are outside the model and can create differences between a least-cost planning pathway and realized investment \citep{Falchetta2021}.

\section{Conclusion}\label{sec_conclusion}

This paper develops a multistage stochastic generation expansion model for power systems under demand and hydrological uncertainty and applies it to Ethiopia and Zimbabwe over 2025--2050. Existing generation expansion studies for sub-Saharan Africa rely mainly on deterministic formulations and provide limited treatment of how sequential demand and hydrological uncertainty affect long-term investment decisions. To address this gap, we formulate the planning problem as a multistage stochastic generation expansion model in which investment decisions adapt to information revealed over time. The SDDP formulation allows future investment commitments to adjust as uncertainty is revealed, subject to technology-specific construction lead times, retirement of inherited capacity, and country-specific technology and resource constraints. The analysis finds a substantial value from preserving this investment flexibility. Relative to the deterministic expected value benchmark, adaptive planning greatly reduces expected unserved energy in both countries. The value of adaptive planning also exceeds the additional value of perfect information by a wide margin. This distinction indicates that the principal gain in the benchmark comparison comes from the ability to revise future investment decisions as conditions evolve, with full foresight providing a smaller additional benefit once such flexibility is available. The associated monetary value remains sensitive to the assumed value of lost load and should be interpreted as a stochastic programming benchmark under the adopted shortage valuation. The one-at-a-time decomposition further shows that hydrological uncertainty has the larger effect on gross resource cost in both applications.

The country results demonstrate how the same planning framework produces different investment responses when initial fleets and resource conditions differ. Ethiopia starts from a large hydropower base and develops a progressively diversified generation portfolio as demand increases. Wind and solar account for substantial additions, even though hydropower remains the largest single source of generation in 2050. The portfolio responds significantly to renewable investment costs, with lower wind and solar costs shifting expansion toward wind and reducing the resource cost of meeting future demand. Hydrological uncertainty remains an important influence on this pathway through its effect on annual hydropower availability. Zimbabwe begins from a hydro-thermal system in which much of the inherited coal fleet is scheduled to retire. The baseline retains part of this coal capacity and adds gas and solar, with the pathway sensitive to the assumed economic treatment of the inherited coal fleet. These differences illustrate the importance of evaluating adaptive expansion against the technological and resource conditions of the individual system than expecting a common investment pattern across countries.

The Zimbabwe application also provides evidence on the interaction between carbon policy and generation investment. Carbon taxes reduce emissions with moderate increases in gross resource cost. At approximately matched expected emissions, the annual emissions cap produces higher net resource cost and expected unserved energy than the corresponding carbon tax in the modeled system. The annual ceiling constrains emissions separately in each year, which limits the system's ability to adjust emitting generation when hydropower availability is low. Under the tax policy, annual emissions can respond to system conditions at the prevailing carbon price. The differences between the instruments remain modest and do not imply a general ranking of price and quantity policies. The investment environment is equally relevant to the cost of emissions reduction. Combining the higher carbon tax with concessional solar financing lowers gross resource cost below the baseline, and lower renewable investment costs reduce resource cost further. Carbon policy and the financing of replacement capacity operate through distinct channels that jointly shape the cost of decarbonization in the Zimbabwe application.

For long-term planning, the results indicate that the value of investment flexibility depends on the uncertainties facing the system and on the options available when those uncertainties are resolved. Hydrological conditions, demand growth, inherited assets, retirement assumptions, resource availability, and investment costs together influence the capacity pathway that emerges. The annual formulation is intended to examine these long-term investment decisions, and it does not resolve hourly renewable variability, unit commitment, ramping, within-year reservoir operation, or storage cycling. Expected unserved energy should consequently be interpreted as an annual energy-shortage measure, not as a detailed measure of operational reliability. Extensions with seasonal, spatial, and hydrological detail could examine operational and regional interactions beyond the scope of the analysis presented in this paper. The model could also be extended to regional power systems by clustering systems according to their resources and available potential.

\section*{Data and code availability}

The data and code to reproduce the results are openly available in the
\href{https://github.com/AregaGetaneh/GEP_Ethiopia_and_Zimbabwe_using_SDDP}{project repository}. The repository contains the hydrological parameter data calibration, initial generation fleet, model implementation, model results, computational settings, and instructions for reproducing the analyses in the paper.

\section*{Acknowledgments}

We are grateful for the two anonymous reviewers for their feedback and comments, which substantially helped us improve the paper. This paper received support from the RE-INTEGRATE project, funded by the European Union's Horizon Europe Research and Innovation Programme under Grant Agreement No. 101118217.

\bibliographystyle{elsarticle-harv}
	\bibliography{references}

\begin{thebibliography}{57}
\expandafter\ifx\csname natexlab\endcsname\relax\def\natexlab#1{#1}\fi
\providecommand{\url}[1]{\texttt{#1}}
\providecommand{\href}[2]{#2}
\providecommand{\path}[1]{#1}
\providecommand{\DOIprefix}{doi:}
\providecommand{\ArXivprefix}{arXiv:}
\providecommand{\URLprefix}{URL: }
\providecommand{\Pubmedprefix}{pmid:}
\providecommand{\doi}[1]{\href{http://dx.doi.org/#1}{\path{#1}}}
\providecommand{\Pubmed}[1]{\href{pmid:#1}{\path{#1}}}
\providecommand{\bibinfo}[2]{#2}
\ifx\xfnm\relax \def\xfnm[#1]{\unskip,\space#1}\fi
\bibitem[{Bazilian et~al.(2012)Bazilian, Nussbaumer, Rogner, Brew-Hammond,
  Foster, Pachauri et~al.}]{Bazilian2012}
\bibinfo{author}{Bazilian, M.}, \bibinfo{author}{Nussbaumer, P.},
  \bibinfo{author}{Rogner, H.}, \bibinfo{author}{Brew-Hammond, A.},
  \bibinfo{author}{Foster, V.}, \bibinfo{author}{Pachauri, S.}, et~al.,
  \bibinfo{year}{2012}.
\newblock \bibinfo{title}{Energy access scenarios to 2030 for the power sector
  in sub-{S}aharan {A}frica}.
\newblock \bibinfo{journal}{Utilities Policy} \bibinfo{volume}{20},
  \bibinfo{pages}{1--16}.
\bibitem[{Bellman(1952)}]{Bellman1952}
\bibinfo{author}{Bellman, R.}, \bibinfo{year}{1952}.
\newblock \bibinfo{title}{On the theory of dynamic programming}.
\newblock \bibinfo{journal}{Proceedings of the National Academy of Sciences}
  \bibinfo{volume}{38}, \bibinfo{pages}{716--719}.
\bibitem[{Birge and Louveaux(1997)}]{Birge1997}
\bibinfo{author}{Birge, J.R.}, \bibinfo{author}{Louveaux, F.},
  \bibinfo{year}{1997}.
\newblock \bibinfo{title}{Introduction to Stochastic Programming}.
\newblock Springer Series in Operations Research. \bibinfo{edition}{1st} ed.,
  \bibinfo{publisher}{Springer-Verlag}, \bibinfo{address}{New York}.
\bibitem[{Bissiri et~al.(2024)Bissiri, Moura, Perez, Figueiredo and
  da~Silva}]{Bissiri2024}
\bibinfo{author}{Bissiri, M.}, \bibinfo{author}{Moura, P.},
  \bibinfo{author}{Perez, R.C.}, \bibinfo{author}{Figueiredo, N.C.},
  \bibinfo{author}{da~Silva, P.P.}, \bibinfo{year}{2024}.
\newblock \bibinfo{title}{Generation capacity expansion planning with
  spatially-resolved electricity demand and increasing variable renewable
  energy supply: Perspectives from power pooling in {West Africa}}.
\newblock \bibinfo{journal}{Applied Energy} \bibinfo{volume}{364},
  \bibinfo{pages}{123115}.
\bibitem[{Blimpo et~al.(2024)Blimpo, Dato, Mukhaya and Odarno}]{Blimpo2024}
\bibinfo{author}{Blimpo, M.P.}, \bibinfo{author}{Dato, P.},
  \bibinfo{author}{Mukhaya, B.}, \bibinfo{author}{Odarno, L.},
  \bibinfo{year}{2024}.
\newblock \bibinfo{title}{Climate change and economic development in {A}frica:
  {A} systematic review of energy transition modeling research}.
\newblock \bibinfo{journal}{Energy Policy} \bibinfo{volume}{187},
  \bibinfo{pages}{114044}.
\bibitem[{Blyth et~al.(2007)Blyth, Bradley, Bunn, Clarke, Wilson and
  Yang}]{Blyth2007}
\bibinfo{author}{Blyth, W.}, \bibinfo{author}{Bradley, R.},
  \bibinfo{author}{Bunn, D.}, \bibinfo{author}{Clarke, C.},
  \bibinfo{author}{Wilson, T.}, \bibinfo{author}{Yang, M.},
  \bibinfo{year}{2007}.
\newblock \bibinfo{title}{Investment risks under uncertain climate change
  policy}.
\newblock \bibinfo{journal}{Energy Policy} \bibinfo{volume}{35},
  \bibinfo{pages}{5766--5773}.
\bibitem[{Botterud and Korp\aa{}s(2007)}]{BotterudKorpas2007}
\bibinfo{author}{Botterud, A.}, \bibinfo{author}{Korp\aa{}s, M.},
  \bibinfo{year}{2007}.
\newblock \bibinfo{title}{A stochastic dynamic model for optimal timing of
  investments in new generation capacity in restructured power systems}.
\newblock \bibinfo{journal}{International Journal of Electrical Power \& Energy
  Systems} \bibinfo{volume}{29}, \bibinfo{pages}{163--174}.
\bibitem[{Cai et~al.(2024)Cai, Lemaire, Medici, Melina, Schwerhoff and
  Thube}]{IMFSSAClimate2024}
\bibinfo{author}{Cai, K.}, \bibinfo{author}{Lemaire, T.},
  \bibinfo{author}{Medici, A.}, \bibinfo{author}{Melina, G.},
  \bibinfo{author}{Schwerhoff, G.}, \bibinfo{author}{Thube, S.D.},
  \bibinfo{year}{2024}.
\newblock \bibinfo{title}{Harnessing Renewables in {S}ub-{S}aharan {A}frica:
  Barriers, Reforms, and Economic Prospects}.
\newblock \bibinfo{type}{Technical Report} \bibinfo{number}{2024/005}.
  International Monetary Fund.
\bibitem[{Carlsson et~al.(2020)Carlsson, Demeke, Martinsson and
  Tesemma}]{Carlsson2020}
\bibinfo{author}{Carlsson, F.}, \bibinfo{author}{Demeke, E.},
  \bibinfo{author}{Martinsson, P.}, \bibinfo{author}{Tesemma, T.},
  \bibinfo{year}{2020}.
\newblock \bibinfo{title}{Cost of power outages for manufacturing firms in
  {Ethiopia}: A stated preference study}.
\newblock \bibinfo{journal}{Energy Economics} \bibinfo{volume}{88},
  \bibinfo{pages}{104753}.
\bibitem[{{Copernicus}(2024)}]{GloFASv5}
\bibinfo{author}{{Copernicus}}, \bibinfo{year}{2024}.
\newblock \bibinfo{title}{River discharge and related historical data from the
  global flood awareness system, version 5.0}.
\newblock \DOIprefix\doi{10.24381/cds.a4fdd6b9}. \bibinfo{note}{dataset
  cems-glofas-historical; system version version\_5\_0; LISFLOOD forced by
  ERA5; consolidated product}.
\bibitem[{Dagnachew et~al.(2023)Dagnachew, Choi and Falchetta}]{Dagnachew2023}
\bibinfo{author}{Dagnachew, A.G.}, \bibinfo{author}{Choi, S.M.},
  \bibinfo{author}{Falchetta, G.}, \bibinfo{year}{2023}.
\newblock \bibinfo{title}{Energy planning in {S}ub-{S}aharan {A}frican
  countries needs to explicitly consider productive uses of electricity}.
\newblock \bibinfo{journal}{Scientific Reports} \bibinfo{volume}{13},
  \bibinfo{pages}{13007}.
\bibitem[{De~Queiroz(2016)}]{deQueiroz2016}
\bibinfo{author}{De~Queiroz, A.R.}, \bibinfo{year}{2016}.
\newblock \bibinfo{title}{Stochastic hydro-thermal scheduling optimization: An
  overview}.
\newblock \bibinfo{journal}{Renewable and Sustainable Energy Reviews}
  \bibinfo{volume}{62}, \bibinfo{pages}{382--395}.
\bibitem[{Dixit and Pindyck(1994)}]{Dixit1994}
\bibinfo{author}{Dixit, A.K.}, \bibinfo{author}{Pindyck, R.S.},
  \bibinfo{year}{1994}.
\newblock \bibinfo{title}{Investment under Uncertainty}.
\newblock \bibinfo{publisher}{Princeton University Press},
  \bibinfo{address}{Princeton, NJ}.
\bibitem[{Dowson and Kapelevich(2021)}]{Dowson2021}
\bibinfo{author}{Dowson, O.}, \bibinfo{author}{Kapelevich, L.},
  \bibinfo{year}{2021}.
\newblock \bibinfo{title}{{SDDP.jl}: A {Julia} package for stochastic dual
  dynamic programming}.
\newblock \bibinfo{journal}{INFORMS Journal on Computing} \bibinfo{volume}{33},
  \bibinfo{pages}{27--33}.
\bibitem[{{Ea Energy Analyses}(2025)}]{EEO2025}
\bibinfo{author}{{Ea Energy Analyses}}, \bibinfo{year}{2025}.
\newblock \bibinfo{title}{{Ethiopian Energy Outlook 2025}}.
\newblock \bibinfo{type}{Technical Report}. Ea Energy Analyses / Danish Energy
  Agency.
\newblock \URLprefix
  \url{https://www.ea-energianalyse.dk/wp-content/uploads/2025/06/Ethiopian-Energy-Outlook-2025.pdf}.
\bibitem[{{EEP}(2025)}]{EEP2025}
\bibinfo{author}{{EEP}}, \bibinfo{year}{2025}.
\newblock \bibinfo{title}{{Ethiopian Electric Power}: generation and
  installed-capacity data}.
\newblock \bibinfo{howpublished}{\url{https://www.eep.com.et/}}.
\newblock \bibinfo{note}{Accessed 2026-08-14}.
\bibitem[{Egli et~al.(2023)Egli, Agutu, Steffen and Schmidt}]{Egli2023}
\bibinfo{author}{Egli, F.}, \bibinfo{author}{Agutu, C.},
  \bibinfo{author}{Steffen, B.}, \bibinfo{author}{Schmidt, T.S.},
  \bibinfo{year}{2023}.
\newblock \bibinfo{title}{The cost of electrifying all households in 40
  {S}ub-{S}aharan {A}frican countries by 2030}.
\newblock \bibinfo{journal}{Nature Communications} \bibinfo{volume}{14},
  \bibinfo{pages}{5066}.
\bibitem[{{Ember}(2025)}]{Ember2025}
\bibinfo{author}{{Ember}}, \bibinfo{year}{2025}.
\newblock \bibinfo{title}{Yearly electricity data}.
\newblock
  \bibinfo{howpublished}{\url{https://ember-energy.org/data/yearly-electricity-data/}}.
\newblock \bibinfo{note}{Accessed 2026-08-17}.
\bibitem[{Falchetta et~al.(2021)Falchetta, Dagnachew, Hof and
  Milne}]{Falchetta2021}
\bibinfo{author}{Falchetta, G.}, \bibinfo{author}{Dagnachew, A.G.},
  \bibinfo{author}{Hof, A.F.}, \bibinfo{author}{Milne, D.J.},
  \bibinfo{year}{2021}.
\newblock \bibinfo{title}{The role of regulatory, market and governance risk
  for electricity access investment in sub-{S}aharan {A}frica}.
\newblock \bibinfo{journal}{Energy for Sustainable Development}
  \bibinfo{volume}{62}, \bibinfo{pages}{136--150}.
\bibitem[{Farquharson et~al.(2018)Farquharson, Jaramillo and
  Samaras}]{Farquharson2018}
\bibinfo{author}{Farquharson, D.}, \bibinfo{author}{Jaramillo, P.},
  \bibinfo{author}{Samaras, C.}, \bibinfo{year}{2018}.
\newblock \bibinfo{title}{Sustainability implications of electricity outages in
  sub-{S}aharan {A}frica}.
\newblock \bibinfo{journal}{Nature Sustainability} \bibinfo{volume}{1},
  \bibinfo{pages}{589--597}.
\bibitem[{Fleten et~al.(2007)Fleten, Maribu and Wangensteen}]{Fleten2004}
\bibinfo{author}{Fleten, S.E.}, \bibinfo{author}{Maribu, K.M.},
  \bibinfo{author}{Wangensteen, I.}, \bibinfo{year}{2007}.
\newblock \bibinfo{title}{Optimal investment strategies in decentralized
  renewable power generation under uncertainty}.
\newblock \bibinfo{journal}{Energy} \bibinfo{volume}{32},
  \bibinfo{pages}{803--815}.
\bibitem[{Fuss et~al.(2008)Fuss, Szolgayov\'{a}, Obersteiner and
  Gusti}]{Fuss2008}
\bibinfo{author}{Fuss, S.}, \bibinfo{author}{Szolgayov\'{a}, J.},
  \bibinfo{author}{Obersteiner, M.}, \bibinfo{author}{Gusti, M.},
  \bibinfo{year}{2008}.
\newblock \bibinfo{title}{Investment under market and climate policy
  uncertainty}.
\newblock \bibinfo{journal}{Applied Energy} \bibinfo{volume}{85},
  \bibinfo{pages}{708--721}.
\bibitem[{Gebremeskel et~al.(2021)Gebremeskel, Ahlgren and
  Beyene}]{Gebremeskel2021}
\bibinfo{author}{Gebremeskel, D.H.}, \bibinfo{author}{Ahlgren, E.O.},
  \bibinfo{author}{Beyene, G.B.}, \bibinfo{year}{2021}.
\newblock \bibinfo{title}{Long-term evolution of energy and electricity demand
  forecasting: The case of {Ethiopia}}.
\newblock \bibinfo{journal}{Energy Strategy Reviews} \bibinfo{volume}{36},
  \bibinfo{pages}{100671}.
\bibitem[{Gebremeskel et~al.(2023)Gebremeskel, Ahlgren and
  Beyene}]{Gebremeskel2023}
\bibinfo{author}{Gebremeskel, D.H.}, \bibinfo{author}{Ahlgren, E.O.},
  \bibinfo{author}{Beyene, G.B.}, \bibinfo{year}{2023}.
\newblock \bibinfo{title}{Long-term electricity supply modelling in the context
  of developing countries: The {OSeMOSYS}-{LEAP} soft-linking approach for
  {Ethiopia}}.
\newblock \bibinfo{journal}{Energy Strategy Reviews} \bibinfo{volume}{45},
  \bibinfo{pages}{101045}.
\bibitem[{Goulder and Parry(2008)}]{GoulderParry2008}
\bibinfo{author}{Goulder, L.H.}, \bibinfo{author}{Parry, I.W.H.},
  \bibinfo{year}{2008}.
\newblock \bibinfo{title}{Instrument choice in environmental policy}.
\newblock \bibinfo{journal}{Review of Environmental Economics and Policy}
  \bibinfo{volume}{2}, \bibinfo{pages}{152--174}.
\bibitem[{{Government of Zimbabwe}(2025)}]{ZWECompact2025}
\bibinfo{author}{{Government of Zimbabwe}}, \bibinfo{year}{2025}.
\newblock \bibinfo{title}{National Energy Compact for the Republic of
  Zimbabwe}.
\newblock \bibinfo{type}{Technical Report}. Mission 300 (World Bank Group and
  African Development Bank).
\bibitem[{Guigues(2017)}]{Guigues2017}
\bibinfo{author}{Guigues, V.}, \bibinfo{year}{2017}.
\newblock \bibinfo{title}{Dual dynamic programming with cut selection:
  Convergence proof and numerical experiments}.
\newblock \bibinfo{journal}{European Journal of Operational Research}
  \bibinfo{volume}{258}, \bibinfo{pages}{47--57}.
\bibitem[{Harrigan et~al.(2020)Harrigan, Zsoter, Alfieri, Prudhomme, Salamon,
  Wetterhall, Barnard, Cloke and Pappenberger}]{Harrigan2020}
\bibinfo{author}{Harrigan, S.}, \bibinfo{author}{Zsoter, E.},
  \bibinfo{author}{Alfieri, L.}, \bibinfo{author}{Prudhomme, C.},
  \bibinfo{author}{Salamon, P.}, \bibinfo{author}{Wetterhall, F.},
  \bibinfo{author}{Barnard, C.}, \bibinfo{author}{Cloke, H.},
  \bibinfo{author}{Pappenberger, F.}, \bibinfo{year}{2020}.
\newblock \bibinfo{title}{{GloFAS-ERA5} operational global river discharge
  reanalysis 1979--present}.
\newblock \bibinfo{journal}{Earth System Science Data} \bibinfo{volume}{12},
  \bibinfo{pages}{2043--2060}.
\newblock \DOIprefix\doi{10.5194/essd-12-2043-2020}.
\bibitem[{He et~al.(2012)He, Wang and Wang}]{HeWangWang2012}
\bibinfo{author}{He, Y.}, \bibinfo{author}{Wang, L.}, \bibinfo{author}{Wang,
  J.}, \bibinfo{year}{2012}.
\newblock \bibinfo{title}{Cap-and-trade vs. carbon taxes: A quantitative
  comparison from a generation expansion planning perspective}.
\newblock \bibinfo{journal}{Computers \& Industrial Engineering}
  \bibinfo{volume}{63}, \bibinfo{pages}{708--716}.
\bibitem[{Hole et~al.(2025)Hole, Philpott and Dowson}]{HolePhilpottDowson2024}
\bibinfo{author}{Hole, J.}, \bibinfo{author}{Philpott, A.B.},
  \bibinfo{author}{Dowson, O.}, \bibinfo{year}{2025}.
\newblock \bibinfo{title}{Capacity planning of renewable energy systems using
  stochastic dual dynamic programming}.
\newblock \bibinfo{journal}{European Journal of Operational Research}
  \bibinfo{volume}{322}, \bibinfo{pages}{573--588}.
\bibitem[{{IEA}(2022)}]{IEA2023Africa}
\bibinfo{author}{{IEA}}, \bibinfo{year}{2022}.
\newblock \bibinfo{title}{{Africa Energy Outlook 2022}}.
\newblock \bibinfo{type}{Technical Report}. International Energy Agency.
  \bibinfo{address}{Paris}.
\newblock \bibinfo{note}{World Energy Outlook Special Report}.
\bibitem[{{IRENA}(2023)}]{IRENA2023}
\bibinfo{author}{{IRENA}}, \bibinfo{year}{2023}.
\newblock \bibinfo{title}{Renewable Power Generation Costs in 2022}.
\newblock \bibinfo{type}{Technical Report}. International Renewable Energy
  Agency. \bibinfo{address}{Abu Dhabi}.
\newblock \URLprefix
  \url{https://www.irena.org/Publications/2023/Aug/Renewable-Power-Generation-Costs-in-2022}.
\bibitem[{Juturu et~al.(2023)Juturu, Conlon, Zhang, Avraam, Siddiqui, Simane
  and Zaitchik}]{Juturu2023}
\bibinfo{author}{Juturu, P.}, \bibinfo{author}{Conlon, T.},
  \bibinfo{author}{Zhang, Y.}, \bibinfo{author}{Avraam, C.},
  \bibinfo{author}{Siddiqui, S.}, \bibinfo{author}{Simane, B.},
  \bibinfo{author}{Zaitchik, B.}, \bibinfo{year}{2023}.
\newblock \bibinfo{title}{Optimal grid expansion under future electricity
  demand for groundwater irrigation in {Ethiopia}}.
\newblock \bibinfo{journal}{Energy for Sustainable Development}
  \bibinfo{volume}{72}, \bibinfo{pages}{351--377}.
\bibitem[{Kolstad(1996)}]{Kolstad1996}
\bibinfo{author}{Kolstad, C.D.}, \bibinfo{year}{1996}.
\newblock \bibinfo{title}{Learning and stock effects in environmental
  regulation: the case of greenhouse gas emissions}.
\newblock \bibinfo{journal}{Journal of Environmental Economics and Management}
  \bibinfo{volume}{31}, \bibinfo{pages}{1--18}.
\bibitem[{Koltsaklis and Dagoumas(2018)}]{KoltsaklisDagoumas2018}
\bibinfo{author}{Koltsaklis, N.E.}, \bibinfo{author}{Dagoumas, A.S.},
  \bibinfo{year}{2018}.
\newblock \bibinfo{title}{State-of-the-art generation expansion planning: A
  review}.
\newblock \bibinfo{journal}{Applied Energy} \bibinfo{volume}{230},
  \bibinfo{pages}{563--589}.
\bibitem[{Mass\'{e} and Gibrat(1957)}]{Masse1959}
\bibinfo{author}{Mass\'{e}, P.}, \bibinfo{author}{Gibrat, R.},
  \bibinfo{year}{1957}.
\newblock \bibinfo{title}{Application of linear programming to investments in
  the electric power industry}.
\newblock \bibinfo{journal}{Management Science} \bibinfo{volume}{3},
  \bibinfo{pages}{149--166}.
\bibitem[{Mekonnen et~al.(2018)Mekonnen, Duan, Rientjes and
  Disse}]{Mekonnen2018}
\bibinfo{author}{Mekonnen, D.F.}, \bibinfo{author}{Duan, Z.},
  \bibinfo{author}{Rientjes, T.}, \bibinfo{author}{Disse, M.},
  \bibinfo{year}{2018}.
\newblock \bibinfo{title}{Analysis of combined and isolated effects of land-use
  and land-cover changes and climate change on the upper {Blue Nile} river
  basin's streamflow}.
\newblock \bibinfo{journal}{Hydrology and Earth System Sciences}
  \bibinfo{volume}{22}, \bibinfo{pages}{6187--6207}.
\newblock \DOIprefix\doi{10.5194/hess-22-6187-2018}.
\bibitem[{Mensah(2024)}]{Mensah2024}
\bibinfo{author}{Mensah, J.T.}, \bibinfo{year}{2024}.
\newblock \bibinfo{title}{Jobs! {E}lectricity shortages and unemployment in
  {Africa}}.
\newblock \bibinfo{journal}{Journal of Development Economics}
  \bibinfo{volume}{167}, \bibinfo{pages}{103231}.
\bibitem[{Murphy and Smeers(2005)}]{Murphy2005}
\bibinfo{author}{Murphy, F.H.}, \bibinfo{author}{Smeers, Y.},
  \bibinfo{year}{2005}.
\newblock \bibinfo{title}{Generation capacity expansion in imperfectly
  competitive restructured electricity markets}.
\newblock \bibinfo{journal}{Operations Research} \bibinfo{volume}{53},
  \bibinfo{pages}{646--661}.
\bibitem[{Pereira and Pinto(1991)}]{Pereira1991}
\bibinfo{author}{Pereira, M.V.F.}, \bibinfo{author}{Pinto, L.M.V.G.},
  \bibinfo{year}{1991}.
\newblock \bibinfo{title}{Multi-stage stochastic optimization applied to energy
  planning}.
\newblock \bibinfo{journal}{Mathematical Programming} \bibinfo{volume}{52},
  \bibinfo{pages}{359--375}.
\bibitem[{Petersen(1973)}]{Petersen1973}
\bibinfo{author}{Petersen, E.R.}, \bibinfo{year}{1973}.
\newblock \bibinfo{title}{A dynamic programming model for the expansion of
  electric power systems}.
\newblock \bibinfo{journal}{Management Science} \bibinfo{volume}{20},
  \bibinfo{pages}{656--664}.
\bibitem[{Pfenninger et~al.(2014)Pfenninger, Hawkes and
  Keirstead}]{Pfenninger2014}
\bibinfo{author}{Pfenninger, S.}, \bibinfo{author}{Hawkes, A.},
  \bibinfo{author}{Keirstead, J.}, \bibinfo{year}{2014}.
\newblock \bibinfo{title}{Energy systems modeling for twenty-first century
  energy challenges}.
\newblock \bibinfo{journal}{Renewable and Sustainable Energy Reviews}
  \bibinfo{volume}{33}, \bibinfo{pages}{74--86}.
\bibitem[{Philpott and Guan(2008)}]{PhilpottGuan2008}
\bibinfo{author}{Philpott, A.B.}, \bibinfo{author}{Guan, Z.},
  \bibinfo{year}{2008}.
\newblock \bibinfo{title}{On the convergence of stochastic dual dynamic
  programming and related methods}.
\newblock \bibinfo{journal}{Operations Research Letters} \bibinfo{volume}{36},
  \bibinfo{pages}{450--455}.
\bibitem[{Schr\"{o}der and Kuckshinrichs(2015)}]{SchroderKuckshinrichs2015}
\bibinfo{author}{Schr\"{o}der, T.}, \bibinfo{author}{Kuckshinrichs, W.},
  \bibinfo{year}{2015}.
\newblock \bibinfo{title}{Value of lost load: {A}n efficient economic indicator
  for power supply security? {A} literature review}.
\newblock \bibinfo{journal}{Frontiers in Energy Research} \bibinfo{volume}{3},
  \bibinfo{pages}{55}.
\bibitem[{Shapiro(2011)}]{Shapiro2011}
\bibinfo{author}{Shapiro, A.}, \bibinfo{year}{2011}.
\newblock \bibinfo{title}{Analysis of stochastic dual dynamic programming
  method}.
\newblock \bibinfo{journal}{European Journal of Operational Research}
  \bibinfo{volume}{209}, \bibinfo{pages}{63--72}.
\bibitem[{Shapiro et~al.(2009)Shapiro, Dentcheva and
  Ruszczy{\'n}ski}]{ShapiroDentchevaRuszczynski2009}
\bibinfo{author}{Shapiro, A.}, \bibinfo{author}{Dentcheva, D.},
  \bibinfo{author}{Ruszczy{\'n}ski, A.}, \bibinfo{year}{2009}.
\newblock \bibinfo{title}{Lectures on Stochastic Programming: Modeling and
  Theory}.
\newblock \bibinfo{publisher}{Society for Industrial and Applied Mathematics
  (SIAM)}, \bibinfo{address}{Philadelphia, PA}.
\bibitem[{Spalding-Fecher et~al.(2017)Spalding-Fecher, Joyce and
  Winkler}]{Spalding2017}
\bibinfo{author}{Spalding-Fecher, R.}, \bibinfo{author}{Joyce, B.},
  \bibinfo{author}{Winkler, H.}, \bibinfo{year}{2017}.
\newblock \bibinfo{title}{Climate change and hydropower in the {Southern
  African Power Pool} and {Zambezi River Basin}: system-wide impacts and policy
  implications}.
\newblock \bibinfo{journal}{Energy Policy} \bibinfo{volume}{103},
  \bibinfo{pages}{84--97}.
\bibitem[{Steinbuks and Foster(2010)}]{Steinbuks2010}
\bibinfo{author}{Steinbuks, J.}, \bibinfo{author}{Foster, V.},
  \bibinfo{year}{2010}.
\newblock \bibinfo{title}{When do firms generate? evidence on in-house
  electricity supply in {Africa}}.
\newblock \bibinfo{journal}{Energy Economics} \bibinfo{volume}{32},
  \bibinfo{pages}{505--514}.
\bibitem[{Sun et~al.(2025)Sun, Anderson, Wei and Zhang}]{Sun2025}
\bibinfo{author}{Sun, X.}, \bibinfo{author}{Anderson, H.M.},
  \bibinfo{author}{Wei, W.}, \bibinfo{author}{Zhang, X.}, \bibinfo{year}{2025}.
\newblock \bibinfo{title}{Energy transition and climate policy selection with
  stochastic demand: Evidence from {Australian} electricity generation
  expansion planning}.
\newblock \bibinfo{journal}{Energy Economics} \bibinfo{volume}{146},
  \bibinfo{pages}{108397}.
\bibitem[{Tekleab et~al.(2011)Tekleab, Uhlenbrook, Mohamed, Savenije, Temesgen
  and Wenninger}]{Tekleab2020}
\bibinfo{author}{Tekleab, S.}, \bibinfo{author}{Uhlenbrook, S.},
  \bibinfo{author}{Mohamed, Y.}, \bibinfo{author}{Savenije, H.H.G.},
  \bibinfo{author}{Temesgen, M.}, \bibinfo{author}{Wenninger, J.},
  \bibinfo{year}{2011}.
\newblock \bibinfo{title}{Water balance modeling of {Upper Blue Nile}
  catchments using a top-down approach}.
\newblock \bibinfo{journal}{Hydrology and Earth System Sciences}
  \bibinfo{volume}{15}, \bibinfo{pages}{2179--2193}.
\bibitem[{Trotter et~al.(2017)Trotter, McManus and Maconachie}]{Trotter2017}
\bibinfo{author}{Trotter, P.A.}, \bibinfo{author}{McManus, M.C.},
  \bibinfo{author}{Maconachie, R.}, \bibinfo{year}{2017}.
\newblock \bibinfo{title}{Electricity planning and implementation in
  sub-{S}aharan {A}frica: A systematic review}.
\newblock \bibinfo{journal}{Renewable and Sustainable Energy Reviews}
  \bibinfo{volume}{74}, \bibinfo{pages}{1189--1209}.
\bibitem[{Weitzman(1974)}]{Weitzman1974}
\bibinfo{author}{Weitzman, M.L.}, \bibinfo{year}{1974}.
\newblock \bibinfo{title}{Prices vs.\ quantities}.
\newblock \bibinfo{journal}{The Review of Economic Studies}
  \bibinfo{volume}{41}, \bibinfo{pages}{477--491}.
\bibitem[{{World Bank}(2024)}]{WorldBankPower2024}
\bibinfo{author}{{World Bank}}, \bibinfo{year}{2024}.
\newblock \bibinfo{title}{Carbon Pricing in the Power Sector: Role and Design
  for Transitioning Toward Net-Zero Carbon Development}.
\newblock \bibinfo{type}{Technical Report}. World Bank.
  \bibinfo{address}{Washington, DC}.
\newblock \URLprefix \url{https://hdl.handle.net/10986/42091}.
\bibitem[{{World Bank Group} et~al.(2020){World Bank Group}, {ESMAP} and
  {Solargis}}]{GlobalSolarAtlas}
\bibinfo{author}{{World Bank Group}}, \bibinfo{author}{{ESMAP}},
  \bibinfo{author}{{Solargis}}, \bibinfo{year}{2020}.
\newblock \bibinfo{title}{Global solar atlas 2.0: Photovoltaic power potential
  (practical potential, pvout level 1)}.
\newblock \bibinfo{howpublished}{\url{https://globalsolaratlas.info}}.
\newblock \bibinfo{note}{Country factsheets for Ethiopia and Zimbabwe}.
\bibitem[{{Zambezi River Authority}(2023)}]{ZRA}
\bibinfo{author}{{Zambezi River Authority}}, \bibinfo{year}{2023}.
\newblock \bibinfo{title}{Zambezi river hydrology: Long-term mean annual flow
  at {Victoria Falls}}.
\newblock \bibinfo{howpublished}{\url{http://www.zambezira.org}}.
\newblock \bibinfo{note}{Long-term mean annual flow approximately
  1{,}100~m$^3$/s}.
\bibitem[{{Zimbabwe Energy Regulatory Authority}(2024)}]{ZERA2024}
\bibinfo{author}{{Zimbabwe Energy Regulatory Authority}}, \bibinfo{year}{2024}.
\newblock \bibinfo{title}{Annual Report 2024}.
\newblock \bibinfo{type}{Technical Report}. Zimbabwe Energy Regulatory
  Authority (ZERA). \bibinfo{address}{Harare, Zimbabwe}.
\bibitem[{Zou et~al.(2019)Zou, Ahmed and Sun}]{Zou2019}
\bibinfo{author}{Zou, J.}, \bibinfo{author}{Ahmed, S.}, \bibinfo{author}{Sun,
  X.A.}, \bibinfo{year}{2019}.
\newblock \bibinfo{title}{Multistage stochastic unit commitment using
  stochastic dual dynamic integer programming}.
\newblock \bibinfo{journal}{IEEE Transactions on Power Systems}
  \bibinfo{volume}{34}, \bibinfo{pages}{1814--1823}.

\end{thebibliography}

	\appendix

\section{Calibration of demand, hydrology, and costs}\label{app_calibration}

The demand process is calibrated from annual electricity demand for Ethiopia and Zimbabwe over 2000--2023 \citep{Ember2025}. A log-linear trend is first fitted to each country series, followed by a first-order autoregression for deviations from the estimated trend. The country estimates are similar given the length of the annual samples. The stochastic model uses a common persistence parameter and relative innovation scale, while retaining country-specific central demand trajectories. Ethiopia is calibrated to observed domestic electricity consumption and the 2030 electrification target. Zimbabwe is based on electricity delivered into the transmission system, which includes the contribution of net imports in the reference year \citep{ZERA2024}. Table \ref{tab_calibration} presents the demand parameters used in the model.

\begin{table}[htbp]
	\centering
	\caption{Demand process calibration for Ethiopia and Zimbabwe.\tablefootnote{The innovation scale is $\sigma_t=\sigma_0\bar{\phi}_t$. The Zimbabwe reference demand consists of 9,830.4 GWh of domestic generation and 1,251.6 GWh of net imports delivered into the transmission system.}\label{tab_calibration}}
	\begin{threeparttable}
		\small
		\begin{tabular*}{\textwidth}{@{\extracolsep{\fill}}lrr}
			\toprule
			Parameter & Ethiopia & Zimbabwe\\
			\midrule
			2025 reference demand (TWh)          & 25.18 & 11.082\\
			2030 electrification target (TWh)    & 40.0  & --\\
			Central growth rate (\%/year)         & 3.5   & 3.5\\
			High-demand growth rate (\%/year)     & 5.6   & 4.69\\
			Fitted persistence                    & 0.84  & 0.78\\
			Pooled persistence $\rho$              & 0.81  & 0.81\\
			Relative innovation scale $\sigma_0$  & 0.077 & 0.077\\
			\bottomrule
		\end{tabular*}
	\end{threeparttable}
\end{table}

Hydrological uncertainty is calibrated from GloFAS version 5 river-discharge histories for 1980--2024 \citep{GloFASv5}, whose reanalysis framework is described by \citet{Harrigan2020}. Table \ref{tab_hydro_discharge} shows the gauge-consistent grid cell used for each basin, the empirical cutoffs that define the low-, normal-, and high-inflow conditions, and the long-run flow levels.

\begin{table}[htbp]
	\centering
	\caption{Hydrological calibration and long-run flow comparison.\tablefootnote{The GloFAS grid cells are located at 11.225$^\circ$N, 34.975$^\circ$E for Ethiopia and $-17.925^\circ$, 25.825$^\circ$ for Zimbabwe. External long-run flow estimates are from \citet{Mekonnen2018} for the Blue Nile and \citet{ZRA} for the Zambezi.}\label{tab_hydro_discharge}}
	\begin{threeparttable}
		\small
		\begin{tabular*}{\textwidth}{@{\extracolsep{\fill}}llrrrrr}
			\toprule
			Country & Basin proxy & $q_{1/3}$ & $q_{2/3}$ & GloFAS mean & External mean & Difference\\
			& & \multicolumn{4}{c}{(m$^3$/s)} & (\%)\\
			\midrule
			Ethiopia  & Blue Nile at El Diem & 1,681.4 & 1,929.1 & 1,793.8 & 1,606.6 & +12\\
			Zimbabwe  & Zambezi at Victoria Falls & 992.7   & 1,299.7 & 1,159.5 & 1,100.0 & +5\\
			\bottomrule
		\end{tabular*}
	\end{threeparttable}
\end{table}

Within each hydrological condition, the availability multiplier is calculated from mean annual inflow and normalized using the stationary distribution\footnote{The stationary distribution of a Markov chain is the long-run share of years the process spends in each state once it has settled down, irrespective of the state in which it started.} of the estimated Markov chain. The normalization imposes $\sum_h \pi_h\mu_h=1$, aligning long-run expected hydrological availability with the central hydropower capacity factor. Transition probabilities are estimated separately for the two basins from the 44 consecutive year transitions in the historical sample. Table \ref{tab_transitions} combines the normalized multipliers, stationary probabilities, and transition matrices. Raw transition counts are given in parentheses to retain the underlying empirical information without duplicating the matrix in a separate panel.

\begin{table}[htbp]
	\centering
	\caption{Hydrological state parameters and transition probabilities.\tablefootnote{$L$, $N$, and $H$ denote low-, normal-, and high-inflow conditions. Transition probabilities are obtained by row-normalizing the empirical transition counts shown in parentheses. Before stationary-distribution normalization, the within-condition inflow multipliers are 0.835/1.003/1.162 for Ethiopia and 0.601/1.005/1.395 for Zimbabwe.}\label{tab_transitions}}
	\begin{threeparttable}
		\small
		\begin{tabular*}{\textwidth}{@{\extracolsep{\fill}}llrrrrr}
			\toprule
			Country & Current condition & $\mu_h$ & $\pi_h$ & $P(L\mid h)$ & $P(N\mid h)$ & $P(H\mid h)$\\
			\midrule
			Ethiopia & Low & 0.84 & 0.348 & 0.500 (7) & 0.214 (3) & 0.286 (4)\\
			& Normal & 1.01 & 0.334 & 0.267 (4) & 0.333 (5) & 0.400 (6)\\
			& High & 1.17 & 0.318 & 0.267 (4) & 0.467 (7) & 0.267 (4)\\
			\midrule
			Zimbabwe & Low & 0.60 & 0.340 & 0.357 (5) & 0.500 (7) & 0.143 (2)\\
			& Normal & 1.01 & 0.325 & 0.467 (7) & 0.133 (2) & 0.400 (6)\\
			& High & 1.40 & 0.335 & 0.200 (3) & 0.333 (5) & 0.467 (7)\\
			\bottomrule
		\end{tabular*}
	\end{threeparttable}
\end{table}

The observed 2024 condition is used to initialize the hydrological state, giving a low-inflow condition for both countries at the beginning of the planning horizon. As a robustness check, the transition probabilities are re-estimated with mild smoothing that adds one-half to every transition count before row-normalizing, giving $P^H_{ij}=(N_{ij}+0.5)/(N_i+1.5)$ with $N_i$ the row total (a Jeffreys prior).\footnote{A Jeffreys prior is a standard uninformative prior in Bayesian estimation. Adding one-half to each count before normalizing is the posterior mean it implies for a three-state chain, which stops a transition absent from the 44-year sample from receiving a probability of exactly zero.} This prevents any rarely observed transition from receiving a probability of exactly zero and limits the influence of the small 44-year sample. The resulting investment outcomes remain close to those obtained from the empirical matrix.
Table \ref{tab_hydro_discharge} also compares the GloFAS basin means with independent long-run estimates. GloFAS gives a somewhat higher mean for each basin, and no level correction is applied, since normalizing the multipliers relative to the long-run basin mean removes a purely multiplicative difference.
Technology availability, investment costs, fixed operating costs, and operational emissions coefficients are consolidated in Table \ref{tab_tech_calibration}. Solar capacity factors are derived from practical photovoltaic potential yields from the Global Solar Atlas \citep{GlobalSolarAtlas}. Hydropower factors and the Ethiopian wind factor are annual planning benchmarks. Hydrological variation in reservoir generation enters separately through the Markov multiplier. The factors assigned to coal, gas, geothermal, and bioenergy represent annual planning level utilization ceilings. Investment cost trajectories follow technology-specific assumptions informed by \citet{IRENA2023}.

\begin{table}[htbp]
	\centering
	\caption{Technology and cost calibration used in the expansion model.\tablefootnote{Solar factors are resource-derived. Hydropower and Ethiopian wind factors are annual planning benchmarks. Coal, gas, geothermal, and bioenergy factors are annual utilization ceilings. Emission coefficients are operational planning coefficients. The bioenergy coefficient follows the net accounting convention used in the model. The continued-operation experiment for coal is treated separately from the baseline cost trajectory. The planning ceiling is the exogenous upper bound on cumulative installed capacity in \eqref{eq_capbound}. It is a long-term planning assumption and not an estimate of technical resource potential. Accelerated access and the high-demand expanded-ceiling case raise the Ethiopian hydropower, wind, and solar bounds to 20, 15, and 15 GW, and the constrained-hydropower case limits cumulative new hydropower construction to 3 GW.}\label{tab_tech_calibration}}
	\begin{threeparttable}
		\small
		\begin{tabular*}{\textwidth}{@{\extracolsep{\fill}}llrrrrrr}
			\toprule
			Country & Technology & Availability & \multicolumn{2}{c}{CAPEX (\$/kW)} & Fixed O\&M & Emissions & Planning\\
			\cmidrule(lr){4-5} & & factor & 2025 & 2050 & (\$/kW-year) & (kgCO$_2$/MWh) & ceiling (GW)\\
			\midrule
			Ethiopia & Hydro       & 0.45  & 1,200 & 1,000 & 30 & 0 & 15.0\\
			& Wind        & 0.34  & 1,300 & 900   & 25 & 0 & 10.0\\
			& Solar PV    & 0.196 & 900   & 400   & 15 & 0 & 10.0\\
			& Geothermal  & 0.88  & 3,000 & 2,500 & 40 & 50 & 5.0\\
			& Bioenergy   & 0.55  & 2,000 & 1,325 & 25 & 230 & 3.0\\
			\midrule
			Zimbabwe & Hydro       & 0.50  & 1,691 & 1,550 & 28 & 0 & 5.0\\
			& Coal        & 0.55  & 1,800 & --    & 40 & 924 & 1.9\\
			& Solar PV    & 0.203 & 900   & 400   & 14 & 0 & 5.0\\
			& Gas         & 0.50  & 1,000 & 850   & 20 & 630 & 3.0\\
			& Bioenergy   & 0.55  & 1,500 & 1,300 & 42 & 230 & 1.0\\
			\bottomrule
		\end{tabular*}
	\end{threeparttable}
\end{table}

The operational emissions coefficients in Table \ref{tab_tech_calibration} are used directly in the generation emissions accounting. Coal and gas coefficients are consistent with standard combustion-based values under representative thermal efficiencies. The bioenergy coefficient is a net planning value, while the geothermal coefficient represents fugitive non-combustion emissions. Hydro, wind, and solar are assigned zero operational emissions. The central value of lost load is 10,000 \$/MWh, with 5,000 and 20,000 \$/MWh used in the sensitivity analysis. This parameter is treated as a planning benchmark. Published values vary considerably across systems and consumer groups \citep{SchroderKuckshinrichs2015}. Evidence from sub-Saharan Africa nevertheless indicates substantial economic costs from unreliable electricity supply \citep{Mensah2024}.

\section{Convergence and sensitivity diagnostics}\label{app_convergence}

Algorithm \ref{alg_sddp} summarizes the SDDP training procedure for the multistage formulation in \eqref{eq_model}. Each iteration consists of a forward pass that samples demand and hydrological realizations and records the resulting trial states, followed by a backward pass that evaluates the stage problems at those states. Dual solutions from the backward pass generate affine cuts that refine the approximation of the continuation value in \eqref{eq_cuts_refined}. Training is monitored using a statistical stopping diagnostic based on the relative difference between the SDDP lower bound and an upper confidence bound obtained from repeated policy evaluation.

\begin{algorithm}[H]
	\small
	\caption{SDDP training procedure for the multistage GEP.}\label{alg_sddp}
	\begin{algorithmic}[1]
		\Require Model \eqref{eq_model}, transition matrix $P^{H}$, Gauss-Hermite demand nodes, and stopping parameters $(\epsilon,\underline{N},\overline{N},P,W)$
		\State Initialize $\mathcal{K}_t^{(i)}\gets\varnothing$ for each stage $t$ and hydrological condition $i$
		\State Set stability counter $c\gets0$
		\For{$n=1,2,\ldots,\overline{N}$}
		\Statex \textbf{Forward pass}
		\State Sample demand innovations $\{\varepsilon_t\}$ and hydrological conditions $\{h_t\}$ with $h_{t+1}\sim P^{H}(\cdot\mid h_t)$
		\State Initialize $x_1$
		\For{$t=1,\ldots,T$}
		\State Solve the stage-$t$ problem at $(x_t,h_t)$ using the current continuation value cuts
		\State Record the trial state and update $x_{t+1}$
		\EndFor
		\Statex \textbf{Backward pass}
		\For{$t=T,\ldots,1$}
		\For{each hydrological condition $i$}
		\For{each Gauss--Hermite demand node}
		\State Solve the stage $t$ problem at the trial state and record the relevant dual multipliers
		\EndFor
		\State Construct the probability-weighted cut $(\beta_{tk},\pi_{tk})$ and append it to $\mathcal{K}_t^{(i)}$
		\EndFor
		\EndFor
		\State Update the lower bound $\underline{V}$ from the first-stage problem
		\If{$n\ge\underline{N}$ and $(n-\underline{N})\bmod P=0$}
		\State Evaluate the policy on the fixed in-sample path set
		\State Construct the one-sided $95\%$ upper confidence bound $\overline{V}$
		\State Compute $g\gets(\overline{V}-\underline{V})/|\underline{V}|$
		\State Update $c\gets c+1$ if $g\le\epsilon$, and set $c\gets0$ otherwise
		\EndIf
		\If{$c\ge W$}
		\State \textbf{break}
		\EndIf
		\EndFor
		\Ensure Cut sets $\{\mathcal{K}_t^{(i)}\}$ defining the trained policy
	\end{algorithmic}
\end{algorithm}

The model implementation uses SDDP.jl under Julia 1.10.4 with Gurobi 13.0.2. Table \ref{tab_convergence} summarizes the training and evaluation settings together with the baseline convergence diagnostics. Policy evaluation for the stopping rule uses a fixed set of paths generated from the same five-point Gauss-Hermite probability model employed in training. The sample state is restored after each evaluation where the stopping calculation does not alter subsequent forward-pass exploration. The one-sided confidence bound is used only for convergence monitoring and is not interpreted as a formal sequential coverage guarantee. Reusing the same evaluation paths reduces Monte Carlo variation in the stopping statistic.

\begin{table}[htbp]
	\centering
	\caption{SDDP training, evaluation, and baseline convergence diagnostics.\tablefootnote{The lower bound and out-of-sample (OOS) cost are given as total expected cost. The final column is the relative difference between the OOS estimate and the training lower bound. Out-of-sample evaluation uses 1,000 independently generated paths under each of five random seeds.}\label{tab_convergence}}
	\begin{threeparttable}
		\small
		
		\begin{tabular*}{\textwidth}{@{\extracolsep{\fill}}lr}
			\multicolumn{2}{l}{\textit{(a) Training and evaluation settings}}\\
			\toprule
			Setting & Value\\
			\midrule
			Demand quadrature                         & 5 Gauss-Hermite nodes\\
			Maximum training iterations               & 700 (1500 for one scenario)\\
			First stopping evaluation                 & Iteration 30\\
			Evaluation frequency                      & Every 5 iterations\\
			In-sample paths per stopping evaluation   & 500\\
			Relative stopping threshold $\epsilon$    & 3\%\\
			Required consecutive evaluations $W$      & 5\\
			Upper confidence bound                    & Mean $+1.645$ standard errors\\
			Out-of-sample evaluation                  & 5,000 paths $(1,000\times5)$\\
			\bottomrule
		\end{tabular*}
		
		\medskip
		
		\begin{tabular*}{\textwidth}{@{\extracolsep{\fill}}lrrrr}
			\multicolumn{5}{l}{\textit{(b) Baseline convergence diagnostics}}\\
			\toprule
			Country & Lower bound (\$B) & OOS cost (\$B) & 95\% CI half-width (\$B) & Difference (\%)\\
			\midrule
			Ethiopia & 15.42 & 15.76 & 0.06 & 2.2\\
			Zimbabwe & 9.70  & 9.85  & 0.04 & 1.5\\
			\bottomrule
		\end{tabular*}
		
	\end{threeparttable}
\end{table}

Across the simulation runs, the stopping diagnostic is small at termination and shows similar behavior for the two applications. The relative difference between the statistical upper bound and the SDDP lower bound lies between about 1\% and 3\% at termination. Every specification terminates on the stopping diagnostic. The combined carbon-tax and concessional-solar-financing specification approaches the tolerance most slowly and is allocated a larger iteration budget, terminating after 410 iterations with a terminal difference of about 2\%. All trained policies are evaluated on 5,000 independently generated out-of-sample paths.

The baseline results in panel (b) of Table \ref{tab_convergence} provide an additional comparison between the training bound and independent policy evaluation. Out-of-sample total expected cost exceeds the training lower bound by about 2.2\% for Ethiopia and 1.5\% for Zimbabwe. The corresponding confidence intervals are narrow relative to the estimated costs, supporting the numerical stability of the baseline evaluations without treating the difference as a certified optimality gap. Additional sensitivity checks examine two numerical features of the stochastic representation. The first increases the demand quadrature from five to nine Gauss-Hermite nodes. The second changes hydrological persistence while preserving the stationary distribution. Table \ref{tab_robustness} shows that these modifications have limited effects on gross resource cost, technology capacity, emissions, and expected unserved energy. Gross resource cost changes by about 1\% or less in both countries, and the main structure of the expansion pathway is preserved.

\begin{table}[htbp]
	\centering
	\caption{Sensitivity of baseline outcomes to demand discretization and hydrological persistence.\label{tab_robustness}}
		\small
		\begin{tabular*}{\textwidth}{@{\extracolsep{\fill}}lrrrrr}
			\toprule
			Specification & Cost & Hydro & Solar & Emissions & EUE\\
			& (\$B) & (GW) & (GW) & (Mt) & (TWh)\\
			\midrule
			\multicolumn{6}{l}{\textit{Ethiopia}}\\
			Baseline, 5 GH nodes           & 16.87 & 9.7 & 8.3 & 0.70 & 0.05\\
			9 GH nodes                     & 16.86 & 9.8 & 8.6 & 0.61 & 0.06\\
			Lower hydrological persistence & 16.82 & 9.8 & 8.7 & 0.75 & 0.07\\
			Higher hydrological persistence& 16.85 & 9.7 & 8.8 & 0.75 & 0.06\\
			\midrule
			\multicolumn{6}{l}{\textit{Zimbabwe}}\\
			Baseline, 5 GH nodes           & 10.21 & 1.8 & 4.7 & 2.80 & 0.04\\
			9 GH nodes                     & 10.19 & 1.7 & 4.7 & 2.70 & 0.04\\
			Lower hydrological persistence & 10.22 & 1.8 & 4.7 & 2.71 & 0.04\\
			Higher hydrological persistence& 10.26 & 1.9 & 4.6 & 2.72 & 0.04\\
			\bottomrule
		\end{tabular*}

\end{table}

\section{Additional results}\label{app_supp}

This section presents supporting results for the valuation of adaptive planning, the decomposition of uncertainty, and selected scenario assumptions discussed in Section \ref{sec_results}. They complement the main text by separating the effects of individual modeling assumptions without repeating the full scenario analysis. Table \ref{tab_vollvss} shows the value of adaptive planning under alternative values of lost load. The investment and operating policies are unchanged across these calculations. Only the monetary valuation of unserved energy is varied. The avoided expected unserved energy is constant, while the estimated monetary value of adaptive planning changes with the assumed shortage cost. This distinction separates the physical benefit of adaptation from its monetary valuation.

\begin{table}[htbp]
	\centering
	\caption{Value of adaptive planning under alternative values of lost load.\label{tab_vollvss}}
		\small
		\begin{tabular*}{\textwidth}{@{\extracolsep{\fill}}lrrrr}
			\toprule & Avoided EUE & \multicolumn{3}{c}{Value of adaptive planning (\$B)}\\
			\cmidrule(lr){3-5}
			Country & (TWh) & \$5,000/MWh & \$10,000/MWh & \$20,000/MWh\\
			\midrule
			Ethiopia & 78.05 & 179.39 & 362.16 & 727.68\\
			Zimbabwe & 15.88 & 39.74  & 80.26  & 161.31\\
			\bottomrule
		\end{tabular*}
\end{table}

The decomposition in Table \ref{tab_uncdecomp} isolates the contribution of demand and hydrological uncertainty. Each specification retains one source of uncertainty while fixing the other at its central trajectory and normal-inflow condition. Hydrological uncertainty has the larger effect on gross resource cost in both countries, with the difference more pronounced in Ethiopia. The joint specification combines both sources and corresponds to the baseline stochastic formulation used in the main analysis.

\begin{table}[htbp]
	\centering
	\caption{Decomposition of demand and hydrological uncertainty.\tablefootnote{The deterministic specification fixes demand at its central trajectory and hydrology at the normal-inflow condition. The demand-only and hydrology-only specifications retain the indicated source of uncertainty. Total expected cost equals gross resource cost less terminal salvage plus the discounted shortage cost.}\label{tab_uncdecomp}}
	\begin{threeparttable}
		\small
		\begin{tabular*}{\textwidth}{@{\extracolsep{\fill}}llrrrr}
			\toprule
			Country & Uncertainty & Gross resource & Shortage & Salvage & Total expected\\
			& represented & cost (\$B) & cost (\$B) & credit (\$B) & cost (\$B)\\
			\midrule
			Ethiopia & Deterministic   & 13.39 & 0.01 & 0.97 & 12.43\\
			& Demand only     & 14.29 & 0.12 & 1.09 & 13.32\\
			& Hydrology only  & 15.88 & 0.00 & 1.17 & 14.71\\
			& Joint           & 16.87 & 0.16 & 1.28 & 15.76\\
			\midrule
			Zimbabwe & Deterministic   & 8.64  & 0.01 & 0.52 & 8.13\\
			& Demand only     & 9.16  & 0.09 & 0.52 & 8.73\\
			& Hydrology only  & 9.42  & 0.02 & 0.43 & 9.00\\
			& Joint           & 10.21 & 0.16 & 0.52 & 9.85\\
			\bottomrule
		\end{tabular*}
	\end{threeparttable}
\end{table}

The Ethiopia experiments in panel (a) of Table \ref{tab_deconf} distinguish faster near-term electrification from sustained long-run demand growth and from the effect of generation resource limits. Accelerated-access requires substantially more wind and solar capacity than the baseline. Under the high-demand trajectory, expanded resource ceilings allow most of the additional requirement to be met through wind and solar investment. Retaining the baseline ceilings changes the composition of the response, increasing reliance on hydropower and geothermal and raising expected unserved energy. The comparison shows that the cost of higher demand depends not only on its level, but also on the generation options available to accommodate it.

For Zimbabwe, panel (b) examines how the treatment of inherited coal affects the expansion pathway. Changes in the timing and implementation of the scheduled retirement floor have only small effects on cost, emissions, and remaining coal capacity. A significantly different pathway emerges when inherited coal incurs a recurring continued-operation cost. Under that specification, most remaining coal capacity is retired, gas investment increases, and emissions fall at a higher gross resource cost. The comparison separates the effect of the prescribed retirement schedule from the economic assumptions constraining operation of the inherited fleet.

\begin{table}[htbp]
	\centering
	\caption{Demand and resource-limit sensitivity for Ethiopia and coal-retirement sensitivity for Zimbabwe.\tablefootnote{Cost denotes gross resource cost. Capacity values refer to 2050. Emissions are average annual operational CO$_2$ emissions unless identified as 2050 values. In panel (a), the expanded-ceiling case relaxes the baseline generation resource limits under the high-demand trajectory. In panel (b), the continued-operation specification assigns a recurring cost of \$600/kW-year to operating inherited coal capacity.}\label{tab_deconf}}
	\begin{threeparttable}
		\footnotesize
		\setlength{\tabcolsep}{3.5pt}
		
		\begin{tabular*}{\textwidth}{@{\extracolsep{\fill}}p{4.0cm}rrrrrrrr}
			\multicolumn{9}{l}{\textit{(a) Ethiopia demand and resource-limit sensitivity}}\\
			\toprule
			Scenario & Cost & Hydro & Wind & Solar & Geothermal & Bioenergy & Emissions & EUE\\
			& (\$B) & (GW) & (GW) & (GW) & (GW) & (GW) & (Mt) & (TWh)\\
			\midrule
			Baseline & 16.87 & 9.7 & 6.5 & 8.3 & 0.2 & 2.6 & 0.70 & 0.05\\
			Accelerated access, 50 TWh by 2030 & 23.02 & 9.7 & 9.7 & 12.1 & 0.2 & 2.7 & 0.81 & 0.04\\
			High demand, expanded ceilings & 24.75 & 9.7 & 9.2 & 13.2 & 3.1 & 2.7 & 1.10 & 0.06\\
			High demand, baseline ceilings & 25.72 & 12.2 & 5.8 & 8.3 & 4.6 & 2.5 & 1.10 & 0.13\\
			\bottomrule
		\end{tabular*}
		
		\medskip
		
		\begin{tabular*}{\textwidth}{@{\extracolsep{\fill}}p{5.5cm}rrrrr}
			\multicolumn{6}{l}{\textit{(b) Zimbabwe coal-retirement sensitivity}}\\
			\toprule
			Scenario & Cost & Average emissions & 2050 emissions & Coal & Gas\\
			& (\$B) & (Mt) & (Mt) & (GW) & (GW)\\
			\midrule
			Baseline retirement floor & 10.21 & 2.80 & 5.31 & 0.7 & 2.1\\
			Schedule only & 10.25 & 2.90 & 5.22 & 0.8 & 2.1\\
			Delayed retirement floor & 10.19 & 2.94 & 5.17 & 0.7 & 2.1\\
			Retirement floor + continued-operation cost & 11.54 & 1.41 & 3.73 & 0.1 & 2.7\\
			\bottomrule
		\end{tabular*}
		
	\end{threeparttable}
\end{table}
Tables \ref{tab_eth_snapshot} and \ref{tab_zwe_policy} present the principal scenarios used in the main Results section.

\end{document}